\documentclass[aps,prd]{revtex4}
\usepackage{mathrsfs}
\usepackage{tikz}
\usetikzlibrary{arrows.meta, positioning, decorations.pathmorphing}
\usepackage{amsmath}
\usepackage{txfonts}
\newcommand{\be}{\begin{equation}}
\newcommand{\ee}{\end{equation}}
\newcommand{\ba}{\begin{eqnarray}}
\newcommand{\ea}{\end{eqnarray}}

\newcommand{\bo}{\raise-1mm\hbox{\Large$\Box$}}
\begin{document}
%\draft
%%%%%%%%%%%%%%%%%%%%%%%%%%%%%

\title{A Few Comments on Classical Electrodynamics}

\author{Kaushik Ghosh\footnote{E-mail ghosh{\_}kaushik06@yahoo.co.in}}
\affiliation{AL - 123, Sector - 2, Kolkata - 700091, India}

\maketitle

\section*{Abstract}

In this article we will discuss a few aspects of the spacetime
description of matter and fields. 
In Section:1 we will discuss the completeness of real numbers in the context
of an alternate definition of the straight line as a geometric continuum. According to this definition,
points are not regarded as the basic constituents of a line segment and a line segment is considered to be a fundamental
geometric object. This definition is in particular suitable to coordinatize different points
on the straight line preserving the order properties of real numbers. Geometrically fundamental nature of line segments
are required in physical theories like the string theory.  
We will discuss the cardinality of rational numbers in the later half of Section:1. We will first discuss
what we do in an actual process of counting and define functions well-defined on the set of all
positive integers. We will follow an alternate approach
that depends on the Hausdorff topology of real numbers to demonstrate that 
the set of positive rationals can have a greater cardinality than the set 
of positive integers. This approach is more consistent with an actual act of counting used in statistical mechanics.
This article indicates that the \textit{axiom of choice} 
is a better technique to prove theorems that use second-countability. This is important for the metrization theorems
and physics of spacetime. In Section:2 we will discuss an improved proof of the Poisson's equation.
We will show that the self energy of a point charge can be zero in the potential approach
to evaluate it. In Section:3 we will discuss a few aspects of the equivalence of
the Schwarzschild coordinates and the Kruskal-Szekeres coordinates.
In Section:4 we will make a few comments on general physics including the special theory of relativity and hydrodynamics.

\vspace{0.5cm}

{\bf MSC:} 51P05, 26A03, 03E25, 03E10, 58A05, 81T30, 83C56

\newpage

\section*{1.1 Introduction to Section:1}

In this section, we will discuss an alternate definition of geometric continuum like the 
straight line that is more consistent with the linear continuum structure of real numbers [1,2,3].
Here a continuum need not to be compact [2]. Here, the term straight line refers to the one dimensional
line in Euclidean geometry that can be coordinatized by the real numbers. 
The term infinite straight line corresponds to the extended real line.
We will find that points need not to be the basic constituents of a line segment and a line segment 
can be considered to be a fundamental geometric object. This is consistent with microscopic
physical theories where the elementary particles are considered to be extended objects
without having any finer structure. One such theory is the string theory. 
In the conventional definition, where a straight line is considered to be a collection of points, we have to break 
the linear continuum structure of real numbers to coordinatize different points on the 
straight line. The linear continuum structure of the real numbers is required to
define concepts like finite open intervals and finite limits in general in the set of real
numbers. The alternate definition of straight line proposed in this article is
more consistent with the linear continuum structure of real numbers.
We need not break the linear continuum structure of the real numbers to coordinatize
the straight line. We will reformulate the completeness axiom
of real numbers that is more consistent with the new definition of straight line.

We can generalize the above aspect to higher dimensional geometric continua.
A geometric continuum of dimension $n$ is not just a collection of geometric continua of dimension $n - 1$
and is a fundamental geometric object.
This is important for the foundations of set theory and topology.
We note that the intersection of two line segments like 
$PQ$ and $QR$ can be a point $Q$ which is different from either set
when we consider a line segment to be a fundamental geometric object. 
We also face problem to construct a topology. Considering the straight line only,
we have two kinds of fundamental objects in it: geometrical points and line segments, and
we have to construct open sets to have a convenient topology in the straight line. 
We solve this problem by assuming that the geometric continua of dimension one and zero 
form a universal set with a line segment being a geometric continuum of dimension one 
and a point being a geometric continuum of dimension zero.
We can introduce familiar topologies in this universal set.
The new definition of straight line as a geometric continuum is naturally expressed as a 
connected topological space in the topology given by geometric open intervals on the straight line.
This topology coincides with a topology derived from a geometrical simple order relation present
between different points on the straight line.
This topology satisfy the Hausdorff and normality conditions [2] due to the fundamental
nature of line segments and coincides with the standard topology of real numbers [2] through coordinatization.  
Note that to begin with there are now two kinds of 'points' present in the topological space. In this article, we refer
Hausdorff condition to mean separability of geometric points while normality indicates
separability of disjoint line segments. It is more natural to rename normality condition 
as also the Hausdorff condition in the straight line and we will discuss this aspect in a future article.

In subsection 1.3, we will discuss a new approach to find the cardinal property of the set of rational
numbers. We denote the set of all positive integers: $\{ 1,2,3...\}$ 
by $Z_{+}$ and the set of all non-negative integers: $\{ 0,1,2,3...\}$ by $\omega$.
To begin, we consider the two dimensional
Euclidean plane. We will emphasize on what we do in an actual process of counting 
and discuss an improved description of injective functions 
defined on limit ordinals [3] like $Z_{+}$. 
The discussions are similar to those applied to show that $S^1$ can not be coordinatized by a single $R^1$
using the stereographic projection from its north pole [2,19]. In this article, two sets are said to be isomorphic
if there is a bijection between the two. We first note that the equivalence class $[Z_{+}]$ of sets
isomorphic to $Z_{+}$ do not possess the order property $O1$ mentioned in the next subsection. 
We now consider two sets of points on the $XY$ -plane given by:
$W_{1} = \{ (n,0)| ~ n \in Z_{+} \}$ and $W_{2} = \{ (-n,0)| ~ n \in Z_{+} \}$. The sets gives
two different representations of $Z_{+}$ and each is bijectively related with $Z_{+}$. They are also bijectively related
with each other through a reflection about the origin. 
Thus $Z_{+}, W_{1}$ and $W_{2}$ have the same cardinality by the $Schr\ddot{o}der-Bernstein$ theorem [3]
and are countable [2]. The discrete set $W = W_{1} \bigcup W_{2}$ is the union of two 
mutually disjoint discrete sets $W_{1}$ and $W_{2}$ and is 
expected to have a greater cardinality than each of them unless we ignore the different 
bijections used to construct $W_{1}$ and $W_{2}$ from $Z_{+}$. The
$Schr\ddot{o}der-Bernstein$ theorem is applicable even to uncountable sets like connected intervals 
of non-zero length from the set of real numbers $R$. There exist
bijections between any pair of these sets due to the order property $O1$ an they are uncountable. 
$W, W_{1}$ and $W_{1}$ do not possess the order property $O1$ and we may have to construct a domain larger than each of 
$W_{1}$ and $W_{2}$ to define an injective function on $W$.
We next consider a convergent sequence of positive rationals $\{ r_{n} \}, ~ n \in Z_{+}$, isomorphic
to $Z_{+}$ and having the same cardinality as that of $Z_{+}$ by the $Schr\ddot{o}der-Bernstein$ theorem [3].
This is the case with $\{ r_{n} = 3/2 + 1/n \}, ~ n \in Z_{+}$. 
The set $G = \{ 1/2 \} \bigcup \{ r_{n} = 3/2 + 1/n \}, ~ n \in Z_{+}$, is
larger than $\{ r_{n} =  3/2 + 1/n \}, ~ n \in Z_{+},$ and is expected to have a greater cardinality than $Z_{+}$.
We will discuss in details the problems that arise when 
we try to construct injective functions from $W$ or $G$ to $Z_{+}$.
We will demonstrate that the set of positive rationals $Q_{+}$ 
have a greater cardinality than the set of positive integers $Z_{+}$. 
We will use Theorem \textbf{1}-7.1. [2], the Hausdorff topology of the real numbers 
and convergent sequences of rational numbers isomorphic to $Z_{+}$  
to demonstrate this. It follows that the set of rationals has a greater cardinality than the set of integers. 
This can give us a departure from the \textit{continuum hypothesis} [2,3].
We find that the \textit{axiom of choice} [2,3] can be a better technique to prove
theorems that use second-countability [2]. This is important for the metrization theorems and is relevant
to quantum gravity. The conclusions of this section are consistent with many 
results from statistical physics that use convergent functions to compare
the cardinalities of sets like $Z_{+}$ and their Cartesian products.
An important example is lattice vibrations [21]. Dispersion relations indicate that an infinite 
lattice with $p$ atoms in every primitive cell contain $p$ times more degrees of freedom in the form of optical branches than
an infinite lattice containing only one atom in its primitive cell.   
Entropy and specific heat of an infinite monatomic lattice with one longitudinal acoustical mode and
two transverse acoustical modes is three times greater than that of the same lattice
with only one longitudinal acoustical mode [21]. This indicates that the cardinality of
the set $Z \times Z \times Z$ is greater than that of $Z$, the set of all
integers including zero. Later, we will discuss another example
that is useful to find the thermodynamic variables for black body radiation.

\section*{1.2 Geometric Continuum, Coordinatization and Hausdorff Condition in the Straight Line}

In this section, we will confine our attention on the straight line 
which is a geometric continuum of one dimension. We will give a precise 
definition of geometric continuum shortly. Henceforth, we will denote the straight line by the 
symbol $E^1$ although we have not yet introduced any metric on the straight line. 
We now define a topology in $E^1$. For any two distinct points $a, b$ on $E^1$, we can define
a geometrical simple order relation between them that asserts that $a < b$ if $b$ lies at the right of $a$.
We choose a basis element as $(a,b)$: the open interval defined as the entire line interval between $a$ and $b$ excluding themselves
with $a < b$ in the geometrical simple order relation. We get the null set when $a$ and $b$ coincide. 
We get a topology on $E^1$ by adding the null set to the basis elements obtained above.
We denote this topology by $\mathcal{U}$. 
The line segment $ab$, also denoted by $[a,b]$, is the the entire line interval between $a$ and $b$ that includes $a$ and $b$.  
We can generate different basis elements by considering $a,b$ to be a pair of indices that take values from
the same index set $K$. $K$ should be large enough to give all possible geometric open intervals on $E^1$.  
It is not a trivial issue that $(a,b)$ can be expressed as an open interval from the set of real numbers $R$
which fixes $K = R = (-\infty,\infty)$. This is
done through the completeness axiom of Dedekind which states that for any point on the number axis, a straight
line, there corresponds a real number from $R$ [4,5]. The real numbers include the rational and
irrational numbers. We can then coordinatize different points
on $E^1$ when we choose a suitable point on $E^1$ to be represented by zero
and a particular point to be represented by one.
The completeness axiom and the algebraic properties of real numbers as a field leads us to 
introduce the Euclidean metric on $E^1$ [2]. The distance between two points with coordinates 
$x,y$ is given by $|x - y|$. The set of real numbers $R$ form an algebraic field that 
has an order relation $<$ that satisfy the following order properties [2]:
 
\vspace{0.5cm}

\noindent{O1. The order relation $ < $ has the least upper bound property. Every nonempty subset
$X$ of $R$ that is bounded above has a least upper bound.}

\vspace{0.5cm}

\noindent{O2. If x $<$ y, then there exists an element z such that x $<$ z and z $<$ y, where $x,y,z \in R$.}

\vspace{0.5cm}

\noindent{Existence of coordinatization indicates that 
$E^1$ can possess the above simple order relation present in $R$ with the properties [2]:}

\vspace{0.5cm}

\noindent{O1. For any nonempty $A \subset {E^1}$, there is a least upper bound for $x(A)$ when $x(A)$
is bounded above.}

\vspace{0.5cm}

\noindent{O2. If $x(a) < x(c)$ there is a point $b$ between $a$ and $c$ such that $x(a) < x(b)$ and $x(b) < x(c)$}.

\vspace{0.5cm}

\noindent{Here $x(a)$ denotes the coordinate of the point $a \in {E^1}$ and $x(A)$ 
includes the coordinate for any point that belongs to the subset $A$ of $E^1$. 
The order property of $R$ introduce an order topology in $E^1$. In $E^1$, 
the topology $\mathcal{U}$ can be now given by $(x(a), x(b))$. This topology and the order
topology induced by $R$ through coordinatization coincide when we choose $x(a) < x(b)$ [2]. 
This topology also coincides with the standard topology of $R$ given by the open sets
$(x,y), ~ x,y \in R$ and $x < y$. Note that we can use $(x(a), x(b))$ to denote open balls in the 
Euclidean metric topology through the midpoint property of Euclidean metric in the straight line. 
$(x(a),x(b))$ gives us an open ball centered at the midpoint of the interval.
The linear order property $O2$ brings us closer to an alternate definition of $E^1$ as a 
geometric continuum through the following axiom [1]:}

\vspace{0.5cm}

AXIOM 1.2-1. \textit{Points are not the basic constituents of a straight line. A line segment is not
just a collection of points and is a fundamental geometric object.}

\vspace{0.5cm}

\noindent{The above axiom may be apparently counter intuitive since two straight lines intersect at a point.
However, whatever may be the cardinality, a collection of zero-extension points can not give an extended 
and homogeneous object like a one dimensional straight line. The same is obviously valid for
line segments. Also a homogeneous collection of
points give us a single point since the extension of a point is zero. We can illustrate the last aspect in the following way:}

\vspace{0.5cm}

\noindent{We consider a homogeneous linear array of marbles touching each
other. If we now shrink the volume of every marble keeping them in
contact (so that the array is always homogeneous), we will get a single point when the volume of every marble 
is made to vanish.}

\vspace{0.5cm}

Axiom 1.2-1 is consistent with the order property O1. We do not need adjacent points to construct a line.
The notion of adjacent points was introduced not to have gaps in the straight line
when we consider it to be merely a collection of points [6]. 
This notion contradicts O1 [6] and can not give an extended object like a line segment without spoiling homogeneity. 
Axiom 1.2-1 allows us to impose the order property $O2$ on the geometrical simple
order relation present between any pair of distinct points in $E^1$ mentioned at the beginning of 
the present section where $a < b$ if $b$ lies at the right of $a$.
Every pair of different points on $E^1$ are now separated by an open line interval,
and we always have a point in between any such pair. 
Thus the geometrical simple order relation satisfies the order property O2.
It is not difficult to impose the order property $O1$ on the geometrical simple order relation 
present in $E^1$. We can introduce the symbol $\infty$ to represent unboundedness of $E^1$
in this order relation. $E^1$ then includes all points $e$ with $-\infty < e < \infty$. This 
indicates that $E^1$ is unbounded in both directions similar to the real numbers. 
We can impose $O1$ to any proper subset of $E^1$ bounded in the geometric order relation.
However, we note that $E^1$ is no longer merely a collection of points. We will elaborate
this aspect shortly.

A few physical aspects of the defining Axiom 1.2-1 of geometric continuum is mentioned in [1].
Here we note that Axiom 1.2-1 is required in the string theory
of theoretical physics [7]. In string theory, point particles are replaced by
one dimensional strings as the most elementary building blocks of matter and radiation.
We can not assume a string to be a collection of points, in
which case it will not be a fundamental object and we will have to hypothesize a new interaction
that will hold the points forming a string. Since string theory involves high energy physics
of quantum gravity, it is unlikely that we will have such an interaction that will always bind
the points forming a string at the high energy scale of quantum gravity. Similar discussions hold
if we assume that elementary particles like electrons or quarks are not point particles 
and have a radius at least greater than their Schwarzschild radius. 
The Schwarzschild radius of an electron is approximately $1.35 \times 10^{-57} m$.
Note that black holes do not have baryonic or leptonic numbers by the no-hair theorems [8], and
it is unlikely that an electron has a leptonic number as well as a radius smaller than its 
Schwarzschild radius.

We can generalize Axiom 1.2-1 to higher dimensional geometric continua.
A geometric continuum of dimension $n$ is not just a collection of geometric continua of dimension $n - 1$
and is a fundamental geometric object.
This is important for the foundations of set theory and topology.
We note that the intersection of two line segments $[a,b]$ and
$[b,c]$ is a point $\{ b \}$ which is different from either set
when we consider a line segment to be a fundamental geometric object. 
We also face problems to construct a topology.
We now have two kinds of fundamental objects in $E^1$: geometric points and line segments.
We want to use open line intervals to construct a simple topology in $E^1$ [9]. 
We remove this problem by adapting the naive definition of sets:
\textit{sets are collections of objects which are closed under the operations of set theoretic
union and intersection}. This definition allows us to introduce the null set. 
Thus, we assume that geometric continua of different dimensions form 
a universal set with a point being a geometric continuum of dimension zero. 
We represent this universal set by the symbol $\mathscr{M}$.
We confine our attention to only $E^1$ which is a subset of $\mathscr{M}$ and try to 
construct a topology for itself. It is now possible to define set theoretic operations
involving line segments and points since they are subsets of the same universal set. We 
construct a topology similar to $\mathcal{U}$ by defining $(a,b)$ as 
$[a,b] - {\{ a \}} - {\{ b \}} = [a,b]\bigcap{\{ a \}^c}\bigcap{\{ b \}^c}$.
We represent this topology by a different symbol $\mathcal{V}$, since $(a,b)$ is no longer
a mere collection of points. $(a,b)$ is not also a collection of line segments and points. It is introduced to give
a topology. We can not introduce $(a,b)$ as a set without defining $\mathscr{M}$. 
We need not to introduce $(a,b)$ to do pure geometry. 
We can also use geometric points and line segments to give a topology [9].    
We now derive a topology from the geometrical simple order relation. The basis elements of this topology consist of
(1) the null set and the straight line, (2) for each point $a$ on the line, the whole line interval at the left of $a$
excluding $a$, (3) open intervals $(a,b)$ for every pair of distinct points $a,b$ on the line with $a < b$
and (4) for every point $b$ on the line, the whole line interval at the right of $b$ excluding $b$.
Again we note that, we have used the geometric order relation between any pair of distinct points on $E^1$ 
to construct open sets that are not merely collections of points. 
We denote this topology by $\mathcal{D}$.
In this topology, we can not separate the straight line into two mutually disjoint open sets.
Thus, Axiom 1.2-1 is consistent with the formal definition of connected topological spaces in the 
topology $\mathcal{D}$.  
This topology coincides with the topology of geometric open intervals $\mathcal{V}$.
We can represent $E^1$ as $(-\infty,\infty)$ in the topology $\mathcal{V}$.
In $\mathcal{V}$ also, we can not separate $E^1$ into two disjoint open sets $A$ and $B$ such that 
${E^1} = A \bigcup B$. We can consider the above conclusions as topological definition of the straight line as a geometric
continuum that follows from Axiom 1.2-1.

We now proceed to construct a representation of the index set $K$
that is consistent with Axiom 1.2-1.
With Axiom 1.2-1 as the definition of a line segment, we can reformulate the axiom 
of completeness of the real numbers as follows:

\vspace{0.5cm}

AXIOM 1.2-2. \textit{Any segment on the straight line $E^1$ can be ascribed a real
number greater than zero. The number can be either rational or irrational. We call this number
to be the length of the segment. A point has zero length.}

\vspace{0.5cm}

Note that, Axiom 1.2-2 refers to segments in $E^{1}$ and is not
stated in terms of points. We assign the same length to $[a,b], (a,b], [a,b)$ and $(a,b)$. 
The length of a line segment is uniquely determined provided we choose a given line
segment to be of unit length. We now choose a suitable point on $E^{1}$ as the origin and coordinatize
any point on $E^{1}$ according to the length of the line segment joining the two
with a particular line segment chosen to be of unit length. Points on the
right side of the origin are assigned positive coordinates while those on the left have negative coordinates.
With both $E^1$ and $R$ being unbounded in their respective order relations,
we can assume $K = R$, where $K$ is the index set mentioned before. We recover the standard topology of $R$ when 
$a,b$ in $(a,b)$ are replaced by the coordinates of $a$ and $b$ given by $x(a)$ and $x(b)$ respectively.  
$(x(a),x(b))$ now represents the open line interval $(a,b)$ on $E^1$.
As before, we can consider it to be an open ball in the Euclidean metric topology.
However, we no longer consider it to be merely a collection of geometric points.
Thus, we can assume that the topology $\mathcal{D}$
derived from the geometric order relation present in $E^1$, the topology $\mathcal{V}$ of geometric open
intervals in $E^1$, the order topology in $R$ and the standard topology in $R$ coincide through 
Axioms 1.2-1 and 1.2-2. Axiom 1.2-1 and the order property $O1$ of $R$ give us a homeomorphism
between $E^{1}$ and $R$ through coordinatization. However, in this article we will continue to 
use different symbols for them to distinguish between geometry and algebra respectively.

It easily follows from Axiom 1.2-1 that $E^1$ is Hausdorff $(T_{2})$ and normal $(T_{4})$ in the topology $\mathcal{V}$ given
by the geometric open intervals $(a,b)$. In this article, we refer
Housdorff condition to mean separability of geometric points by disjoint open sets while normality indicates
separability of disjoint line segments by disjoint open sets.
Since line segments are fundamental objects, \textit{every} pair of different points on $E^1$ are always
separated by an open line interval. They are contained within an open line interval
and we can construct two disjoint open sets separating the pair   
by deleting any point within the line segment joining the two.  
The same is valid for two disjoint line segments $PQ$ and $RS$ with $P < Q < R < S$ in
the geometric order relation. $PQ$ and $RS$ are closed sets and are contained 
within an open line interval. The points $Q$ and $R$ are always separated by an open line
interval and we can construct two disjoint open sets separating $PQ$ and $RS$   
by deleting any point in the line segment joining $Q$ and $R$. 
Thus, $E^1$ is Hausdorff and normal in the topology $\mathcal{V}$.
However, it would be more appropriate to say that $E^1$ is a Hausdorff space if we
consider the fundamental nature of geometric points and line segments. We do not follow this approach
for the present article. In a coordinatization scheme, the Housdorff and normality conditions are ensured by
the completeness of real numbers and the order property O2 of $R$.
Given a pair of different points $a$ and $c$ of $E^1$ and a coordinatization where $x(a) < x(c)$, 
$O2$ indicates that for any point $b$ in between $a$ and $c$ 
there exists a number $x(b)$ that coordinatize $b$ such that $x(a) < x(b) < x(c)$.  
$(x(a_{-}),x(b)), (x(b),x(c_{+}))$ with ${x(a_{-}) < x(a)}$ and $x(c) < x(c_{+})$ are a pair of 
open sets that separate $a$ and $c$ respectively. Similar arguments can be
used to establish normality condition by considering the coordinates $x(Q)$ and $x(R)$ of $PQ$
and $RS$. This is expected since $(a,b)$ and $(x(a),x(b))$ give the same topology as mentioned before. 
However, the Hausdorff condition in the topology $\mathcal{U}$ does not remain valid 
for every pair of points if we regard the straight line to be merely a 
collection of points. Look at the discussions given below Axiom 2-1 with reference to [6]. 
If we assume the existence of adjacent points $e_{1},e_{2}$ on $E^1$, the sets $(u_{1}, e_{2})$
and $(e_{1}, u_{2})$ with $u_{1} < e_{1} < e_{2} < u_{2}$, that separate $e_{1}$ and $e_{2}$ 
no longer remain open in $\mathcal{U}$. There is no open set containing $e_{1}$ that is contained in $(u_{1}, e_{2})$. Similar
aspect remains valid for $e_{2}$ and $(e_{1}, u_{2})$. 
With this definition, $E^1$ is Hausdorff only in the discrete topology. 
Lastly, we note that $x(a), x(b)$ in $(x(a),x(b))$ can be rational or irrational.
We can not use the basis of open balls with rational radii centered at points with
rational coordinates to give a Hausdorff topology in $E^1$. This is due to the existence of irrationals like 
$\sqrt{2}, \pi$. The points with irrational coordinates should belong to one or other of these basis elements and 
hence can not be separated from every point with rational coordinate by open sets obtained from this covering.
This is important for differential geometry.

\section*{1.3. Cardinality of the Rationals}

In this subsection, we will discuss a new approach to find the cardinal property of the set of rationals.
We will emphasize on what we do in an actual process of counting as in the case of
statistical mechanics to show that the set of positive rationals $Q_{+}$ 
have a greater cardinality than the set of positive integers $Z_{+}$. 
We will use the Hausdorff topology $\mathcal{V}$ of $E^1$
which is same as the standard topology of $R$ through coordinatization. We will call it the standard
topology of $E^1$ in the following. We express $R$ as $(- \infty, \infty)$ with the understanding
that the limit $x \rightarrow \infty$ means $x$ can be increased indefinitely without any upper bound.
It follows that $R$ is an uncountable set and $E^1$ contains an uncountable number of line
segments. The terminology used in this section mostly follow [2].
We use the symbol $|A|$ to express the cardinality of the set $A$. 
The cardinality of a set gives a measure of the amount of elements present in the set. 
In general, the cardinality of a set may not be given by ordinary positive integers. 
The cardinality of a finite set is given by a positive integer and the set is considered
to be countable. We consider the set of positive integers $Z_{+}$ to be countable. 
Any element of $Z_{+}$ can give us the cardinality of a discrete set containing the same
number of elements. $Z_{+}$ is considered to be the largest countable set [2]. Any set
with a greater cardinality is uncountable [2]. We do not include zero in $Z_{+}$ because
zero decides only the requirement of counting and is not an outcome of an actual act of counting. 
We now state Theorem \textbf{1}-7.1., [2]:

\vspace{0.5cm}

THEOREM 1.3-1. \textit{Let $B$ be a nonempty set. Then the following are equivalent:}

\vspace{0.2cm}

\noindent{(1) \textit{There is a surjective function $f: {Z_{+}} \rightarrow B$.}}

\noindent{(2) \textit{There is an injective function $g: B \rightarrow {Z_{+}}$.}}

\noindent{(3) \textit{$B$ is countable.}}

\vspace{0.3cm}

\noindent{The proof can be found in [2]. Isomorphic sets have the same cardinality by the $Schr\ddot{o}der-Bernstein$ theorem [3].
We now demonstrate that the set of rational numbers has a greater cardinality than the set of
integers.}

We first discuss the following comments regarding functions defined on $Z_{+}$. 
These comments are consistent with the inductive
character of $Z_{+}$ and also with the characteristics expected of an actual act of counting. 
We denote the set of first $N$ positive integers $\{1,2,3,..,N \}$ by $Z_{N}$, the set
of first $N$ positive odd integers $\{1,3,5,..,2N - 1 \}$ by $O_{N}$ and the set of  
first $N$ positive even integers $\{2,4,6,..,2N \}$ by $E_{N}$. 
We first note that $|Z_{N}^{q}| = {|Z_{N}|^q}$ for bounded values of
$N$, where $Z_{N}^{q}$ is the $q$ -th order Cartesian product of $Z_{N}$. 
Picking the odd integers from a given $Z_{N}$ is not an injective function on $Z_{N}$
for any $N \in Z_{+}$ apart from $N = 1$. Similar situation remains valid if we want to pick
any proper subset of $Z_{+}$ from $Z_{+}$. These functions are similar to the projection operators.
A different problem arises for injective functions like
$2n - 1$ and $2n, ~ n \in Z_{+}$. To define these functions on $Z_{N}$, the range should
be at least $Z_{2N}$ for all $N \in Z_{+}$. Neither the range $Z_{2N}$ nor these functions are 
well-defined in the limit $N \rightarrow \infty$, \textit{i.e}, 
when $N$ is increased indefinitely without any upper bound. 
This is because there can not exist any integer greater than $N$ in the 
limit $N \rightarrow \infty$, \textit{i.e}, when $N$ is increased indefinitely without any upper bound.
We need to consider the limit $N \rightarrow \infty$ due to the inductive character of $Z_{+}$
containing the set of all positive integers. We also note that $Z_{+}$ being the largest set
of positive integers, we can not propose the existence of a set of positive integers larger than $Z_{+}$.
Thus, functions like $2n$ and $2n - 1$ are not defined on $Z_{+}$ to $Z_{+}$.
The above discussions remain valid if we try to construct any injective function $f(n)$ on $Z_{+}$ such that 
$f(n)$ is finite for finite values of $n$ but $f(n) > n$ as $n \rightarrow \infty$. 
This is the case with $2n - 1, 2n, {m^n}; ~ m,n \in Z_{+}$ and $m$ is finite.
Such functions are required when we try to construct an inejctive function from a a discrete
set $D$ to $Z_{+}$ where $Z_{+}$ is a proper subset of $D$ or is isomorphic to a 
proper subset of $D$. Note that the induction principle ceases to hold in the limit $n \rightarrow \infty$
because $n + 1$ does not exist in this limit.
Similar aspect also remains valid for real variables. To illustrate, we consider two open intervals of $R$ given by
$(-x, x)$ and $(-2x, 2x)$ with $x > 0$. $(-x, x) \subset (-2x, 2x)$ for finite or bounded values of $x$.
However, in the limit $x \rightarrow \infty$, \textit{i.e}, 
when $x$ is increased indefinitely without any upper bound,
$(-x, x)$ coincides with $R$ and the function $f(x) = 2x$ ceases to exist for $x \rightarrow \infty$. 
Similar arguments are applied to show that $S^1$ can not be coordinatized by a single $R^1$
using the stereographic projection from its north pole. To illustrate, we consider the two dimensional
Euclidean plane. In the stereographic projection to $X$ -axis from the north pole of a unit-radius $S^1$ with its south pole
at $(0,0)$, there exists no real number including the integers on the $X$-axis to coordinatize
the north pole having coordinates $(0,2)$. This is because we can not define a real number
$x + \epsilon, \epsilon > 0$ in the limit $x \rightarrow \infty$. 
This aspect is used to demonstrate that there can not exist a homeomorphism 
between $S^1$ and $R^1$ or any connected finite open interval of $R^1$, the later being
homeomorphic to $R^1$. Thus, in the homeomorphism that maps the open interval $(-1,1)$ to $R^1$, there exists
no real number to assign a coordinate to any point outside the interval $(-1,1)$.
As another illustration, we can confine our attention to the open interval $(-4,4)$ homeomorphic to $R$.
We consider the function $2z: (-4,4) \rightarrow (-4,4)$. 
The function is not defined for $|z| \geq 2$ and the function does not exist in the limit $z \rightarrow \pm 4$.
The condition $|z| < 2$ for the function $2z$ on $(-4,4)$ is analogous to the condition of 
boundedness on $x$ for the function $2x$ on $R$. Similar aspect remains valid 
for functions like ${x^2}: R \rightarrow R$ and ${z^2}: (-4,4) \rightarrow (-4,4)$. 
We conclude that we can not express $Z_{+}$ as ${O_{+}}\bigcup{E_{+}}$ where $O_{+} = \{ x| x = 2n - 1, n \in Z_{+} \}$ 
and $E_{+} = \{ x| x = 2n, n \in Z_{+} \}$ since $O_{+}$ and $E_{+}$ 
do not exist. This is not unexpected since $Z_{N} \neq {O_{N}}\bigcup{E_{N}}$
for any finite $N \in Z_{+}$. In particular, we can not construct an injective function
from $\{ 1,2 \} \times {Z_{+}}$ to $Z_{+}$ by using functions like ${m^n}; ~ m,n \in Z_{+}$ and $m$ is finite.
To illustrate, to define an injective function like $f[(k,n)] = {2^k}{5^n}, ~k = 1,2; ~ n \in Z_{+}$ on
$\{ 1,2 \} \times {Z_{+}}$ the range should be at least $Z_{{2^2}{5^n}}$ with 
the limit $n \rightarrow \infty$. The range and the function do not exist in this limit. 
The above aspects are not considered in some discussions that aim at 
demonstrating that $Z_{+} \times Z_{+}$ and $Q_{+}$ are countable [2,3]. In such demonstrations, functions like 
$2n - 1, 2n$, and ${m^n}$ that are well-defined from $Z_{N}$ to $Z_{+}$ for 
finite values of $N$ are inappropriately extended to $Z_{+}$. 
In passing, we note that we can have an 
injective function from $(-2x, 2x) \in R$ to $(-x, x) \in R$, both of which satisfy the order property $O_{1}$.  
However, $Z_{+} \times Z_{+}$ and $Z_{+}$ are discrete collections of points.
We can introduce an order relation in $Z_{+}$ by using the order relation of $R$ and use
the dictionary order relation in $R \times R$ [2] to introduce an order relation in $Z_{+} \times Z_{+}$.
With these order relations, $Z_{+}$ and $Z_{+} \times Z_{+}$ do not have the order property $O_{1}$.

We now mention the following comments. In an actual act of counting of the number of 
elements in a set $A$, we first separate the elements of $A$ and enumerate them 
in an increasing order using the elements of $Z_{+}$ as $1,2,3...$. 
The maximum integer in the above sequence without any gap gives us the cardinality of $A$. 
Using other members of the equivalence class $[Z_{+}]$ of sets isomorphic to $Z_{+}$
for enumeration is not much meaningful in an actual act of
counting. We can use the elements of the set $S_{+}$ given by: ${s_{n}} = f(n), n \in Z_{+}$;
where $f(n)$ is a well-defined injective function, to
enumerate the elements of $A$ but we have to use the index $n \in Z_{+}$ of $s_{n}$ 
and not $s_{n} \in S_{+}$ itself to express the number of elements present in $A$. 
To illustrate further, we consider two towers of points in the plane given by: 
$H_{1} = (1,m), ~ H_{2} = (2,n), ~ 1 \leq m \leq M, ~ 1 \leq n \leq N; ~ m,n,M,N \in Z_{+}$. We can enumerate
the elements of $H_{1}$ using the elements of the set $S_{+}$. We can enumerate the 
the elements of $H_{2}$ using the elements of $T_{+}$ defined as $T_{+} = - S_{+}$. 
Thus, we can enumerate the elements of $H_{1}$ as: $s_{1}, s_{2}, ..., s_{M}$, and the elements of $H_{2}$ 
as: $-s_{1}, -s_{2}, ..., -s_{N}$. However, the cardinality of $H_{1}$ is $M$, that of $H_{2}$ is $N$ and
the cardinality of $H = H_{1} \bigcup H_{2}$ is given by $M + N \in Z_{+}$. Note that, in this case we have
a mapping from a proper subset of $\{ 1,2 \} \times {Z_{+}}$ to ${Z_{+}}$. $|H|$ may not remain well-defined for
all possible choices of $H_{1}$ and $H_{2}$. For example, we can consider $H_{1}$ to consist
of all points with $m \in Z_{+}$ and take $N = 2$ for $H_{2}$. We can easily construct a bijective 
function: $h : H_{1} \rightarrow Z_{+}$ defined as $h[(1,m)] = m$. This
indicates that $|H_{1}| = |Z_{+}|$, by the $Schr\ddot{o}der-Bernstein$ theorem [2,3]. The cardinality of $H$, given by: 
$|H| = |H_{1}| + |H_{2}| = |Z_{+}| + 2$, does not correspond to any integer in $Z_{+}$
and is greater than $|Z_{+}|$. This conclusion justifies that $H$ contains points in addition
to those present in the countable set $H_{1}$. Similar conclusion remains valid if we try to construct
any injective function $h$ from $H$ to $Z_{+}$. The problems with such constructions
was discussed in the above paragraph.   
In the second example, $H_{1}$ is isomorphic to $Z_{+}$ and should have the same cardinality as that of $Z_{+}$.
$H$ should have a cardinality greater than $H_{1}$ unless $H_{2}$ is null. We conclude that $H$ is uncountable in this case.  
The situation is similar to the cardinality of the set $W$ and $G$ mentioned in the introduction. 
As a further illustration, we can use convergent functions from Cartesian products of $\omega$ or $Z_{+}$ to $R$
to measure the number of points present in such products. This is often done in statistical
mechanics. Here we construct an example. We first construct the function $g(n_{1}, n_{2}, .., n_{q})$
from the finite dimensional Cartesian product ${\omega}^{q}$ to $R$ defined as:

\[
g(n_{1}, n_{2}, .., n_{q}) = \exp(-{\alpha}{\sum}_{i}{n_{i}}) = {\prod_{i}}\exp(-{\alpha}{n_{i}}) 
= {\prod_{i}}{[{\exp(-{\alpha})}]^{n_{i}}}, ~~~ {n_{i}} \in {\omega}, ~ \alpha > 0
\]

\noindent{where the product is over the $q$ copies of $\omega$. We next construct the functional:}

\[
\mathcal{Z} = {\sum}_{n_1}..{\sum}_{n_q}{g(n_{1}, n_{2}, .., n_{q})} , ~~~ {n_{i}} \in \omega 
\]

\noindent{This gives,} 

\[
\mathcal{Z} \rightarrow {({{e^{\alpha}} \over {e^{\alpha} - 1}})^q} 
\]

\noindent{Where we have first considered the partial sum of first $n + 1$ terms of the series
${\sum}_{k}{\exp{(-\alpha k})}, ~ k \in \omega$. We express it as:}

\[
S(n) = {1 \over {1 - r}} - {{r^n \over {1 - r}}} + {r^n}, ~~~~ r = \exp{(- \alpha)}
\]

\noindent{The sum over $\omega$ is obtained by taking the limit $n \rightarrow \infty$.
The functions $r^n:(0,1) \rightarrow R$ converge to $0$ when $n \rightarrow \infty$. 
Thus $\mathcal{Z}$ converges to ${({{e^{\alpha}} \over {e^{\alpha} - 1}})^q}$. 
We also consider the functional:}

\[
E = - {\partial \over {\partial \alpha}}{ln(\mathcal{Z})} = {q \over {e^{\alpha} - 1}}.
\]

\noindent{Both $\mathcal{Z}$ and $E$ are defined as weighted sum over the complete $\omega^{q}$ and gives
a measure of $|{\omega}^q|$ using the convergent function $g$ from ${\omega}^q$ to $R$. 
The above expressions for $\mathcal{Z}$ and $E$ indicate
that $|{\omega}^q| > |\omega|$, consistent with the discussions given in this paragraph. For
$\alpha = {{h \nu} / kT}$ and $q = 2$, $\mathcal{Z}$ gives the partition function of 
the photons for a plane electromagnetic wave with frequency $\nu$, [22]. 
Here $h$ is Planck's constant, $k$ is Boltzmann constant, $T$ is temperature and $q = 2$
accounts for two polarizations of the electromagnetic wave. The internal energy of the photons
is given by: $U = 2 {h \nu} / (e^{\beta h \nu} - 1)$, where $\beta = {1 / kT}$. This expression
is useful to explain black body radiation [22], and indicates that $|{\omega}^2| > |\omega|$.    
In the above expression for $\mathcal{Z}$, we can restrict the sums over all $n_{i}$ apart from $n_{1}$ to nonvanishing finite values.
The resulting expression is larger than the case when all $n_{i}$ apart from ${n_1}$ are vanishing.
This indicates that the cardinality of the sets like $H$ mentioned before is greater than $Z_{+}$.}

It is expected from the above discussions that $Q_{+}$ has a greater cardinality than $Z_{+}$. 
We also note that $Q_{+}$ have the order property $O1$ but $Z_{+}$ do not.
We now prove the following theorem. We follow an approach similar to the proof of Theorem \textbf{3}-6.5., [2].
In the proof of this theorem, we will consider a counting where the elements 
of a set are separated using the Housdorff condition. 
We start by noting that we use a bijective function on $Z_{+}$ to define a 
sequence $\{ x_{n} \} ~(n \in Z_{+})$ isomorphic to $Z_{+}$. Any other bijective function from
$Z_{+}$ to the sequence $\{ x_{n} \}$ can be considered  
to be a permutation on the elements of $Z_{+}$ because the index set of $\{ x_{n} \}$ is $Z_{+}$.

\vspace{0.5cm}

THEOREM 1.3-2. \textit{The set of positive rational numbers and the set of positive real numbers are uncountable.}

\vspace{0.2cm}

\textit{Proof}. We will first construct a set $A = S \bigcup X$, where $S$ is a sequence of
positive rationals which is isomorphic to $Z_{+}$ and $X$ is another set of positive rationals whose
elements are different from $S$. We will consider $X$ to be finite. 
Thereafter, we will show that $|Z_{+}| < |A|$. It then follows that: $|Z_{+}| < |Q_{+}|$.

We consider a nested family of closed intervals of ${E^1}$ given by: 
$[3/10, {1/2}], [33/100, {1/2}], [333/1000, {1/2}],...$. 
We index these intervals as $I_{1}, I_{2}, I_{3},...$, with $I_{n}$ containing $I_{n + 1}$
and $n \in Z_{+}$. We now consider a set of points on $E^1$ with coordinates 
${y_{n}} \in R$, $n \in Z_{+}$. For $n \geq 2$, $y_{n}$ is the midpoint between 
the lower boundaries of $I_{n - 1}$ and $I_{n}$ and has coordinate:  
$y_{n} = ({1 \over 2})[{{x_{1}....x_{n - 1}} \over {10^{n - 1}}} + {{x_{1}....x_{n}} \over {10^{n}}}]
= (1/2)(0.x_{1}....x_{n - 1} + 0.x_{1}....x_{n})$,
where all $x_{i}$ are $3$ and ${y_{1}} = 3/20 = 0.15$. All ${y_{n}}$ are different and rationals. 
The first few are given by ${3/20}, {63/200}, {663/2000}, {6663/20000}$. By the Housdorff property of 
the standard topology of $E^1$ or $R$, 
the lower boundary of every $I_{n}$ can be separated from 
the corresponding (in index) $y_{n}$ by two disjoint open intervals and every $I_{n}$ 
excludes all ${y_{m}}$ with $m \leq n$. We can construct the bijective function $f: Z_{+} \rightarrow {R}$ 
by $f(n) = {y_{n}}, ~ n \in Z_{+}$. Each $I_{n}$ excludes the corresponding point ${y_{n}}$. 
We now consider the closed interval $I_{\aleph} = [{5/12}, {1/2}]$. $1/3$ is the limit of
the sequence $\{ u_{n} = {x_{1}....x_{n} \over {10^n}} \}, ~ n \in Z_{+}$ and ${x_{i}} = 3$ for all $i \in Z_{+}$. 
$I_{\aleph}$ is contained in the intersection of all $I_{n}$, contains the points 
with coordinates ${5/12}$ and ${1/2}$, but none of $f(n) = {y_{n}}$ is contained in $I_{\aleph}$. 
We can replace $\aleph$ by the cardinal that represents the cardinality of $Z_{+}$.
We continue to use $\aleph$ with the note that $\aleph \notin Z_{+}$.
We now consider the set $A = \{ y_{1}, y_{2}, ... \} \bigcup \{ {y_{\aleph}} = 1/2 \}$.
The function $f$, although injective, is not a surjection of $Z_{+}$ to $A$. 
There can not exist such a surjection from $Z_{+}$ to $A$ since $\aleph \notin Z_{+}$.
Any one-to-one surjective (bijective) function $g$ from $Z_{+}$ to $A$ can be considered to be a permutation on the elements of $Z_{+}$
onto the index set of the elements of $A$. This can be easily understood if we compose with $g$ from the left 
the bijective function $h$ from $A$ to the index set of its elements where, 
$h(y_{n}) = {f^{-1}(y_{n})} = n, ~ n \in Z_{+}$ and $h({y_{\aleph}} = 1/2) = \aleph$.  
Such a permutation is possible if the the index set of $A$ would have been $Z_{+}$. 
However, we can not permute the elements of $Z_{+}$
to generate an additional element so that they can correspond to the elements of a set
containing one more element in addition to those present in $Z_{+}$. 
Thus, there can not exist a one-to-one surjective function from $Z_{+}$ to $A$.
As mentioned before, it is important to keep in mind that there exists no injetive 
function $f(n)$ from $Z_{+}$ to $Z_{+}$ such that $f(n)$ is finite for finite values 
of $n$ but $f(n) > n$ as $n \rightarrow \infty$. 
We can consider other examples of $S$ and $I_{n}$. We can take $I_{1} = [1/4, 3/2], ~ z_{1} = 8/5 = 1.6$ and 
$I_{n} = [1/4 ,{1/2 + {1/n}}], ~ z_{n} = {1 \over 2}  + ({1 \over 2})[{1 \over {n - 1}} + {1 \over {n}}]$
for $n \geq 2$ and $n \in Z_{+}$. We take $I_{\aleph} = [1/4,3/8]$.
In this case, the point with coordinate $1/2$ is the limit of the sequence $\{ 1/2 + {1/n} \}, ~ n \in Z_{+}$.
We replace $A$ by $B = \{ z_{1}, z_{2}, ... \} \bigcup \{ z_{\aleph} = 1/4 \}$.
Similar arguments as above show that there can not exist a one-to-one surjective (bijective) function from $Z_{+}$ to $B$.

We now show that there can not exist any surjective function from $Z_{+}$ to $A$. 
To elaborate, let there is a surjective function $p$ from $Z_{+}$
to $A$ which is not injective. Let $Y$ be the subset of $Z_{+}$ that is not mapped injectively
to $A$. Let, $p(Y) = P; ~Y \subset Z_{+}, ~P \subset A$ and $|P| < |Y|$. $P$ is a proper subset of $A$ and 
$Y$ can not be $Z_{+}$ if $p$ has to be surjective. For any such $p$, we can 
always redefine the bijective function $f$ used in the previous paragraph to
another bijective function $f': Z_{+} \rightarrow \{ y_{1}, y_{2}, ... \}$ 
so that $f'(Z_{+} - Y)\bigcap P = \emptyset$. We can not extend $p$ to 
a bijective function from $Z_{+} - Y$ to ${[f'(Y) \bigcup \{ y_{\aleph} \} - P]} \bigcup {[f'(Z_{+} - Y)]}$.
The arguments are similar to those used above to prove the corresponding result for $Z_{+}$ and $A$. We should note that 
${f'(Y) \bigcup \{ y_{\aleph} \} - P}$ is nonempty and is different from ${f'(Z_{+} - Y)}$.  
They have different index sets. We now show that we can not have $Y = Z_{+}$ and $P = A$.
In this case, we can restrict $p$ to construct a bijective function $p'$ from a suitable subset of $Z_{+}$
to $A$. $(p')^{-1}$ gives us an injective function from $A$ to $Z_{+}$. 
We exclude these examples of $p$ by showing that we can not have an injective function from $A$ to $Z_{+}$. 
Every injective function $q$ from $A$ to $Z_{+}$
can be considered to be a bijection $q'$ on the indices of $\{ y_{n} \}, ~ n \in Z_{+}$ and ${y_{\aleph}}$
to the image set of $q$. We can use the bijective function ${h^{-1}}$ on the index set of $A$ and compose it
with $q$ from the right $(= q{h^{-1}})$ to understand this. 
A bijection on the elements of a set can not give a lesser number of images.
Since $\aleph \notin Z_{+}$, it is not possible to have an injective function from $A$ to $Z_{+}$
even if the image set of $q$ coincides with $Z_{+}$. 
Similar conclusion is valid for the second example with $A$ replaced by $B$.
Thus, there can not exist a surjective function from $Z_{+}$ to $A$ or $B$. 
There can exist a surjective function from the set $C$ to the set $D$ even if there may not exist any
bijective function from $C$ to $D$ if both the sets have subsets that
satisfy the order property $O1$. This is easy to show when the bijective functions are homeomorphisms. 
We can take $C = [0,2]$ and $D = [0,1] \bigcup \{ 2 \}$. $Z_{+}, A$ and $B$ do not satisfy the order property $O1$. 
Similarly, we could have an injective function from $A$ to $Z_{+}$ if both $A$ and $Z_{+}$ would have 
satisfied the order property $O1$.

It follows from Theorem 1.3-1 that $A$ and $B$ are uncountable and $|Z_{+}| < |A|$ and $|Z_{+}| < |B|$.
$A$ and $B$ are proper subsets of $Q_{+}$. Thus, there can not exist a surjective function from
$Z_{+}$ to $Q_{+}$ and the later is uncountable. It is obvious that $R_{+}$ is also uncountable. Q.E.D.

An alternate proof of the above theorem using inductive procedure will be given later.

\vspace{0.5 cm}

It is always safer to use the inductive procedure
to resolve confusions regarding $[Z_{+}]$. To illustrate, the set $Z_{+} - \{ 1 \}$
can be obtained inductively from $Z_{N} - \{ 1 \} = \{ 2,3,..,N \}, ~ N \geq 2$. We consider 
$(Z_{+} - \{ 1 \}) \notin [Z_{+}]$. $Z_{+} - \{ 1 \}$ can be considered 
to be the set of all positive integers excluding $1$. 
$Z_{+} - \{ 1 \}$ is countable since any element of $Z_{+} - \{ 1 \}$ can represent 
the possible outcome of an act of counting on a countable set except the case when 
the number of elements to count is one or the set is isomorphic to $Z_{+}$.
We consider the cardinality of $Z_{+} - \{ 1 \}$ to be given by the cardinal number $\varepsilon - 1$,
where $\varepsilon$ is the cardinal number: $\varepsilon = |Z_{+}|$. 
$\varepsilon - 1$ is less than $\varepsilon$ although we can not represent the former by any positive integer.
Similarly, the cardinality of the set of all odd integers or even integers contained in $Z_{+}$
can be given by ${\varepsilon / 2}$. Algebraic operations on cardinal numbers and 
non-negative integers can be interpreted as set theoretic operations on suitable members 
of the equivalence classes whose cardinalities are given by corresponding cardinal numbers
and non-negative integers. The equivalence classes are collections of isomorphic sets. 
Thus, the cardinality of the set of integers $Z$ 
can be given by $2 \varepsilon + 1$. 
We now illustrate the significance of the inductive procedure to define $Z_{+}$ with 
a proof of the following lemma. A different version of this lemma 
is proved in the proof of Theorem 1.3-2.
The proof of this lemma together with Theorem 1.3-1 give us an alternate proof of Theorem 1.3-2.

\vspace{0.5cm}

LEMMA 1.3-1. \textit{There can not exist a surjective function from $Z_{+}$ onto $A = {\{ 1/2 \}} \bigcup {Z_{+}}$.}

\vspace{0.2cm}

\textit{Proof}. Here $1/2$ signifies an element different from all elements of $Z_{+}$.  
Let $S$ be a set whose elements are the members of a sequence $\{ s_{n} \}, n \in Z_{+}, s_{n} > {1/2}$ for
all $n \in Z_{+}$ and $s_{i} \neq s_{j}$ for all $i,j \in Z_{+}$.   
There exists a bijective function $f: Z_{+} \rightarrow S$ given by: $f(N) = s_{N}$.
The collection of all such sets are bijectively related with one another. $Z_{+}$ is 
a particular member of this collection. We denote this collection of isomorphic sets by the class $[S]$.   
This is a subset of the equivalence class of all sets that are isomorphic with $Z_{+}$.
We define $S_{N}$ to be the set $\{ s_{1}, s_{2},.., s_{N} \}$, where
$s_{i} \in S \in [S]$. Let $A_{N} = {\{ 1/2 \}}\bigcup Z_{N}$ and $A = {\{ 1/2 \}}\bigcup Z_{+}$. 
It is obvious that there can not exist a surjective function from
$S_{1}, S_{2}, S_{3}$ onto $A_{1}, A_{2}, A_{3}$ respectively for all $S \in [S]$. This follows since a function can not
assign more than one value to a single element of its domain. Suppose, there does not exist a surjective 
function from $S_{N}$ onto $A_{N}$ for all $S \in [S]$. 
We then show that there can not exist a surjective function from $S_{N + 1}$ onto  
${A_{N + 1}} = {A_{N} \bigcup {\{N + 1}\}}$ for any $S \in [S]$. For, if there exists such a function $f$
for an $S \in [S]$, we can have two cases. We may have 
$f(s_{N + 1}) = N + 1 \in A_{N + 1}$. 
Otherwise, we can always rearrange $S_{N + 1}$ to have an $S'_{N + 1}$ with 
$S' \in [S]$ so that $f(s'_{N + 1}) = N + 1 \in A_{N + 1}$.
This can be done in the following way. Let 
$f(s_{k}) = N + 1 \in A_{N + 1}, ~ s_{k} \in S_{N + 1}, ~k \neq {N + 1}$.
We remove $s_{k}$ from $S_{N + 1}$ and construct $S'_{N}$ where $s'_{i} = s_{i}, ~ 1 \leq i < k$
and $s'_{i} = s_{i + 1}, ~ k \leq i \leq N$. 
We then add $s_{k}$ to $S'_{N}$ as $s'_{N + 1}$ to obtain $S'_{N + 1}$. We supplement further the rest of the terms 
$s_{N + 2}, s_{N + 3}, s_{N + 4}, ...$ of $S$ to $S'_{N + 1}$ to construct $S' \in [S]$.
In either case, we can not extend $f$ to be a surjective
function on the rest of $S_{N + 1} (S'_{N + 1})$ onto the rest of $A_{N + 1}$ since, such a function does not exist
by assumption. But the assumption holds for $N = 1,2,3$. Hence by induction, we can not have a surjective function
from $S (S')$ onto $A$. In the later case, existence of a surjective function
from $S$ onto $A$ will give a surjective function from $S'$ onto $A$ since, a bijective function 
from $S'$ to $S$ followed by a surjective function from $S$ onto $A$ is a surjective function
from $S'$ onto $A$. However, we have shown that the later can not exist. Thus, there can not exist a surjective function
from $S$ onto $A$ in both the cases. 
This also holds for $Z_{+} \in [S]$. Existence of a surjective function
from $Z_{+}$ onto $A$ will give a surjective function from $S$ onto $A$ since, a bijective function 
from $S$ to $Z_{+}$ followed by a surjective function from $Z_{+}$ onto $A$ is a surjective function
from $S$ onto $A$ and the later can not exist. Q.E.D

\vspace{0.5cm}

We can also prove the following lemma using the inductive procedure.

\vspace{0.5cm}

LEMMA 1.3-2. \textit{There can not exist an injective function from $A = {\{ 1/2 \}}\bigcup {Z_{+}}$ to $Z_{+}$.}

\vspace{0.2cm}

\textit{Proof}. The proof is similar to that of Lemma 1.3-1. We construct the set $V$ from the set $S$ used
in the proof of Lemma 1.3-1 by $V = {\{ 1/2 \}}\bigcup {S}$ 
and construct a new class $[V]$. The bijections between different elements of $[S]$ can be extended
to bijections between the corresponding elements of $[V]$ by identifying ${\{ 1/2 \}}$ of every element of $[V]$.   
We define $V_{N}$ to be the set ${\{ 1/2 \}}\bigcup \{ s_{1}, s_{2},.., s_{N} \}$, where
$s_{i} \in S \in [S]$.
We note that we can not have an injective function from $V_{1}, V_{2}, V_{3}$ to $Z_{1}, Z_{2}, Z_{3}$ for
all $V \in [V]$. Suppose, there does not exist an injective 
function from $V_{N}$ to $Z_{N}$ for all $V \in [V]$. 
We then show that there can not exist an injective function from $V_{N + 1}$ to  
$Z_{N + 1}$ for any $V \in [V]$. For, if there exists such a function $f$
for an $V \in [V]$, we can have four possibilities. We exclude two possibilities where
either no element of $V_{N + 1}$ is mapped to $N + 1 \in Z_{N + 1}$ or  
$f(1/2) = N + 1 \in Z_{N + 1}$. In both cases, $f$ has to be injective
from $Z_{N + 1}$ to $Z_{N}$. This is not possible for any $N \in Z_{+}$. 
In the other two cases, we may have $f(v_{N + 1}) = N + 1 \in Z_{N + 1}$. 
Otherwise, we can always rearrange $V_{N + 1}$ to have an $V'_{N + 1}$ with 
$V' \in [V]$ so that $f(v'_{N + 1}) = N + 1 \in Z_{N + 1}$.
The construction of $V'$ is similar to that of $S'$ used to prove Lemma 1.3-1. In either case, we can not extend $f$ to be an injective
function on the rest of $V_{N + 1} (V'_{N + 1})$ to the rest of $Z_{N + 1}$ since, such a function does not exist
by assumption. But the assumption holds for $N = 1,2,3$. Hence by induction, we can not have an injective function
from $V (V')$ to $Z_{+}$. In the later case, existence of an injective function
from $V$ to $Z_{+}$ will give an injective function from $V'$ to $Z_{+}$ since, a bijective function 
from $V'$ to $V$ followed by an injective function from $V$ to $Z_{+}$ is an injective function
from $V'$ to $A$. However, we have shown that the later can not exist. Thus, there can not exist an injective function
from $V$ to $Z_{+}$ in all the four cases. This also holds for $A \in [V]$. Existence of an injective function
from $A$ to $Z_{+}$ will give an injective function from $V$ to $Z_{+}$ since, a bijective function 
from $V$ to $A$ followed by an injective function from $A$ to $Z_{+}$ is an injective function
from $V$ to $Z_{+}$ and the later can not exist. Q.E.D

\vspace{0.5cm}

It is now easy to show:

\vspace{0.5cm}

LEMMA 1.3-3. $E^1$ contains an uncountable number of line segments.

\vspace{0.2cm}

\textit{Proof.} This follows from Axiom 1.2-2 and uncountability of $R$.

\vspace{0.5cm}

As an important consequence of the present section, Theorem 1.3-2 indicates that the covering of $R$ by the collection of 
open balls with rational radii centered at points with rational coordinates no longer gives a countable 
covering of $R$. We can use $(r_{1},r_{2})$ to denote such open balls in the 
Euclidean metric topology through the midpoint property of Euclidean metric in the straight line. 
Here, $r_{1}$, $r_{2}$ are two rational numbers and $(r_{1},r_{2})$ gives us an open ball centered at the midpoint of the interval.
This is important for the Lindel$\ddot{o}$f covering theorem in $R$ which
uses the above covering and we will
have to construct a new countable covering to show that $R$ is a Lindel$\ddot{o}$f space [10].  
We will also have to change the proof of Heine-Borel theorem given in [10] accordingly. This proof
uses the Lindel$\ddot{o}$f covering theorem in $R$.  
We may not be able to assume that a smaller segment of a line segment $[a,b]$ is compact
to show that $[a,b]$ is compact [9]. The method of induction does not work here since a closed interval 
$[a,b] \in {R}$ with $a \neq b$ contains an uncountable number of smaller closed intervals.
We will later try to construct an alternate proof 
of the Heine-Borel theorem by using the \textit{axiom of choice}. 
The present article can be significant for differential geometry and quantum gravity. General 
affine connections including non-metricity are pertinent to construct a theory of quantum gravity 
and can be useful to explain dark energy and inflation [11]. 
The corresponding mathematics relies heavily on second-countability [9,12].
Thus, it requires due attention if the real line itself is not second-countable. 
We can try to use the \textit{axiom of choice} as an alternative to prove theorems that use countable basis [9].
We also note that $R$ need not to be first-countable when the rationals are not countable.

\vspace{0.5cm}

From the set theoretic perspective, the non-integral rationals are as different from the integers as the 
irrationals are compared to the rationals. This is in particular significant for counting. 
It is more appropriate to have two different cardinals to represent the 
cardinal properties of $Z_{+}$ and $Q_{+}$ if cardinality is well-defined
for these sets. We have shown in Theorem 1.3-2 that $Q_{+}$ has a greater
cardinality than $Z_{+}$. Existence of the irrationals indicates that we have the hierarchy
$|Z_{+}| < |Q_{+}| < |R_{+}|$. This gives us a departure from the \textit{continuum hypothesis} [2,3].
The proof of Theorem 1.3-2 indicates that ${\varepsilon} + n > {\varepsilon}, ~ n \in Z_{+}$. 
Theorem 1.3-2 also indicates that the cartesian products $Z_{+}^{n}$ are not countable
and ${\varepsilon}^{n} > {\varepsilon}$.
Thus, ${\varepsilon}$ is similar to the positive integers in these
aspects. We can also state that $2^{\varepsilon} > {\varepsilon}$.
This is good for ${\varepsilon}$ to represent the cardinality of a set since the cardinality of the power set of a set 
is always greater than that of the set according to Cantor's theorem on the cardinality of power sets [3]. 
We also use the term cardinal numbers to denote cardinalities like 
${\varepsilon} \pm n$ and ${\varepsilon}^{n}, ~ n \in Z_{+}$, discussed above.

\newpage

\section*{2. Discussions on Poisson's equation}

In this section, we will consider an improved derivation of the
Poisson's equation for electric field of a static source [13,14]. We
first note that we can not use the usual derivation of integral version of the
divergence theorem if the vector field is not sufficiently well-behaved at any point
inside the region of interest [14]. In one dimension, we
can use the relation: ${{\int_{x_1}^{x_2}}{(df/dx)}dx} = {f(x_2) -
f(x_1)}$, only if $f(x)$ is sufficiently well-behaved for all $x$ between and
including the limits. The expression ${E_x{(\Delta{x})}} =
{E_x{(r = 0)}} + {\Delta{x}} [{{d{E_x}}\over{d{x}}}]_{r = 0}$ is not
well-defined for a point charge at the origin. 
We also note that although $r = 0$ is a removable singularity of the function $({{r^2} \over{r^2}})$, 
it is appropriate not to replace it by $1$ at $r = 0$ to evaluate its radial derivatives  
at the origin. The later are required when we use the spherical polar coordinates to evaluate
the divergence of electric field of a point charge at the origin. 
However we will find from the following
discussions that for a point charge at the origin, we may replace ${\vec{E}{(0)}}$ by the value of $\vec{E}$ at any other point 
within a small neighborhood containing the origin to evaluate the volume 
integral of ${\vec{\nabla}}.{\vec{E}}$ using spherical polar coordinates. In this section, we
will show by using the spherical polar coordinates that an explicit
volume integration of the divergence of electric field of a
point charge gives us the total charge. The charge is at the origin included in the region
of volume integration. Also for a point charge at the origin, volume integral of
the divergence of electric field is vanishing when the volume of
integration does not include the origin. The above results, together with the fact
that the total flux of electric field of a point charge at the
origin over an $S^{2}$ centered at the origin is given by $q \over
{\epsilon_0}$, lead us to the integral version of divergence
theorem for the electric field. We can generalize the theorem to a point charge not at
the origin and have the Poisson's equation for electrostatic
field.

We first consider a point charge at a radius vector $\vec{a}$ on the
$Z$ -axis. In the following discussions we will ultimately consider
the point charge to be at the origin. We use the inverse square law
in spherical polar coordinates for the electric field of a point
charge at the position vector $\vec{a}$. The electric field is
azimuthally symmetric and the azimuthal component is zero.
We integrate the divergence of electric field over a sphere
centred at the origin. The radius vector of the point charge is
$\vec{a}$ and the radius of the sphere is greater than $|\vec{a}|$.

We first consider the integral of the radial derivative term:

\be {{4\pi\epsilon_0}}I_{r} = {\int dv}[{1 \over {r^2}}{\partial
\over {\partial r}}({{{r^2}{{\vec R}.{\hat r}}} \over {R^3}})] \ee

\noindent{Here ${\vec R} = \vec{r} - \vec{a}$ and the radius of the volume of
integration is greater than $|\vec{a}|$.
We considered $\vec{a}$ to be on the $Z$-axis. The measure in the
integral contains a $\sin{\theta}$ term in the spherical polar
coordinate system. Hence, the complete triple integrand is a mixed function of
$(r,\theta)$ and is divergent of degree two at the point charge. Thus, the integral is
well-behaved and the Riemann sum remains independent of our choice
of the point of evaluation of the integrand. In particular, we can
evaluate the integrand in the Riemann integral for a very small nonzero value of
the polar angle and use the rule of total integrals to the above radial integral to have,}

\ba {{4\pi\epsilon_0}}I_{r} & = & {\int{d\Omega}{{r^3 q}\over {[r^2 +
a^2 -2ra{\cos{\theta}}]^{3/2}}}} \\ \nonumber 
& & - {\int{d\Omega}{{r^2 q
(\vec{a}.\hat{r})}\over {[r^2 + a^2 -2ra{\cos{\theta}}]^{3/2}}}} \ea

\noindent{In the limit $a = 0$ the above expression gives the
result $I_{r} = {q \over {{\epsilon}_0}}$.
We now consider the integral of polar angular derivative term.
This integral is also a total integral in the polar angle $\theta$.
The upper limit of the integral gives vanishing contribution. The
integral can take finite value only from the lower limit $\theta =
0$ part. This term and hence the value of the corresponding integral
is given by the following expression:}

\be {{4\pi\epsilon_0}}I_{\theta} = -{2
\pi}{\int_0^{r'}}{(rdr)}{\sin{(\delta \theta)}}({{{{\vec R}.{\hat
\theta}}} \over {R^3}}) \ee

\noindent{Here $r' > a$. In the limit ${\delta \theta} \rightarrow 0$ and $a \rightarrow 0$
the above integral vanishes. There is no azimuthal component of the electric
field when the point charge is on the polar axis.
Hence we have $\int{{{{\vec{\nabla}}_{\vec{r}}}}.{\vec{E}}}dv = {q
\over {{\epsilon}_0}}$ when the point charge is at the origin.}

When the point charge is at the origin
${{4\pi\epsilon_0}}\int{{{{\vec{\nabla}}_{\vec{r}}}}.{\vec{E}}}dv$
is vanishing if the integrating volume does not include the origin.
If we consider any annular region centered at the point charge at origin, the volume integral of
$({{{\vec{\nabla}}_{\vec{r}}}}.{\vec{E}})$ is given by the following
expression:

\be \int{q d\Omega}[({{r_2}\over{r_2}})^3 - ({{r_1}\over{r_1}})^3] =
0 \ee

\noindent{It is easy to show that the volume integral of
$({{{\vec{\nabla}}_{\vec{r}}}}.{\vec{E}})$ is vanishing for any
arbitrary volume not including the point charge at origin. The electric
field is well-behaved at any point apart from the origin and we can use 
Gauss' integral law to any region not including the point charge. 
Thus, we have the well-known Poisson's equation for a point charge at
the origin,}

\be 
{{{\vec{\nabla}}_{\vec{r}}}}. {({{{\hat{r}}}\over{r^2}})} = {4
\pi}{{\delta}^3(\vec{r})}
\ee

\noindent{The above discussions together with the fact that the total flux of
the electric field over a closed surface is ${Q \over {\epsilon_0}}$
allows us to write the corresponding integral version of Gauss's
divergence theorem for the electric field of a point charge at the
origin. We can derive a corresponding expression for the divergence of
electric field of a point charge not at the origin. We follow the
above derivation with ${\vec{r}}$ and ${\vec{a}}$ replaced by
${\vec{R}} = ({{\vec{r}} - {\vec{r}'}})$ and ${\vec{a'}} =
({{\vec{r}''} - {\vec{r}'}})$ respectively and consider the explicit
volume integrals over a sphere centred at $\vec {r}'$. Here $\vec
{r}'$ is the position of the point charge.
We consider the inverse square law to find the flux of electric
field and we have the Poisson's equation for electrostatic field
of a point charge not at the origin,}

\be {{{\vec{\nabla}}_{\vec{r}}}}. {({{{\hat{R}}}\over{R^2}})} = {4
\pi}{{\delta}^3(\vec{R})} \ee

\noindent{We can use the above procedures given in this section to apply the divergence theorem of
Gauss to any field of the form $({{\hat{R}} \over {R^n}})$ with $n < 3$. 
A different derivation is given in [15] where we have to change the integrand in Eq.(1).}

We now consider the no-work law for electric field. For a point
charge source, the electric field is singular at the point charge and we may
not be able to apply the Stoke's theorem. We can consider an annular
region surrounding the point charge with the inner boundary being an
infinitesimally small circle centered at the point charge and apply
Stoke's theorem to the electric field in this region. The work done
for the inner boundary vanishes out of spherical symmetry and we
have the no work law for the outer closed line.

We conclude our discussions on Maxwell's equations for steady
sources with a few discussions on the derivation of curl of
$\vec{B}$ law. In the derivation of $({\vec{\nabla} \times
{\vec{B}}})$ law, we have a 'boundary' term [14]:

\be I_{B} = \int{{{\vec{\nabla}}_{\vec{r}'}}.
[{{{\vec{J}}(\vec{r}')}\over{R}}]}d{v'} \ee

\noindent{We can not use Gauss' divergence theorem directly as ${{\vec
J{(\vec{r}')}}\over{R}}$ is divergent at $\vec{R} = 0$. We break the
integral into two parts:}

\be I_{B} = \int_{\delta v}{{{\vec{\nabla}}_{\vec{r}'}}.
[{{{\vec{J}}(\vec{r}')}\over{R}}]}d{v'} + \int_{V - \delta
v}{{{\vec{\nabla}}_{\vec{r}'}}.
[{{{\vec{J}}(\vec{r}')}\over{R}}]}d{v'} \ee

\noindent{Here $\delta v$ is an arbitrarily small sphere surrounding
$\vec{r}$. For a non-singular current distribution, the first term
vanishes in the limit $R \rightarrow 0$. In the second term
${{\vec{J(\vec{r}')}}\over{R}}$ is well-behaved everywhere within
$V - \delta v$. We can use the divergence theorem to transform the
volume integral into two boundary surface integrals. For the inner
boundary, $\int{{{\vec{J(\vec{r}')}}\over{R}}.{d \vec{s_1}}}$ is
vanishing for a regular current distribution as the radius of
$\delta v$ become arbitrarily small. For the outer boundary,
${\vec{J}}.d{\vec{S_2}}$ is zero as the current should be tangential
at the boundary of source. However this derivation is valid when
the current density is non-singular, and the derivatives of $\vec{J}$ are
sufficiently well-behaved within the source apart from the boundary.}

We will now derive an exact expression for the electrostatic energy.
The expression for the electrostatic energy obtained in this section
agrees with the standard expressions when the sources are not
point charges [14,15]. However we show that the electrostatic self-energy of
a point charge is vanishing instead of being infinite. This is in
contrast to the standard expressions where the electrostatic
self-energy of a point charge is infinite [14,15].

When we start from the potential formulation, the expression of the electrostatic energy, 
expressed in terms of the electrostatic potential, is given by the following expression [14,15,16]:

\be
\xi = {1 \over 2}{\int {\rho(\vec{r}')}{V(\vec{r}')} d{\tau}}
\ee

\noindent{The integral can be taken to be over the complete volume. Here,
$V(\vec{r}')$ is the potential at the point $\vec{r}'$ due to
all the other source elements apart from that at the point $\vec{r}'$ itself. 
We denote such quantities by $V_{R}(\vec{r}')$.
However, the charge density is given by the divergence of electric field due to the complete source. 
We denote such quantities by the suffix $T$. Thus the energy is given by the following expression:}

\be
\xi = {{\epsilon_0} \over 2}{\int {({\vec{\nabla}}.{\vec{E}_{T}}})}{V_{R}} d{\tau}
\ee

\noindent{For a non-singular source density both ${V_{R}}$ and
${\vec{E}_{R}}$ can be replaced by ${V_{T}}$ and ${\vec{E}_{T}}$
respectively. The corresponding fields and the potentials are non-singular and are vanishing in the 
limit when source volume element tends to zero. We can replace $V_{R}$ by $V_{T}$ in
Eq.(10) and we can also apply the Gauss'integral law to 
the vector field $(V_{T}{{\vec{E}_{T}}})$.
We then have the conventional expression for electrostatic energy expressed in terms of the electric field:}

\be
\xi = {{\epsilon_0} \over 2}{\int ({{\vec{E}_{T}}.{\vec{E}_{T}}})d{\tau}}
\ee

\noindent{Here ${\vec{E}_{T}}$ is the complete field at the point $(x,y,z)$.
However, the situation is different when we have point charges. 
In these cases, $V_{R}$ is finite at the point charges. On the other
hand $V_{T}$ is divergent at the point charges and
the value of the energy becomes different if
we replace $V_{R}$ by $V_{T}$ in Eq.(10). In fact
replacing $V_{R}$ by $V_{T}$ in Eq.(10) for point charges,
gives us the usual divergent value for electrostatic self energy
of point charges provided we can apply the Gauss' integral law
to fields divergent as ${1 \over {r^3}}$. We have found at the beginning of this section, that the application of Gauss' law
to fields divergent as ${1 \over {r^3}}$ is a non-trivial issue.}

In this section we evaluate the electrostatic energy from Eq.(10) without replacing $V_{R}$ by $V_{T}$.
When we have an isolated point charge, $V_{R}$ is zero at the
point charge but is finite for any other point. Let us consider the
point charge to be at the origin. The product $(V_{R}{\vec{E}_T})$
varies as ${1 \over {r^3}}$ but is zero at the point charge. As we
have discussed earlier, we can not apply the divergence theorem to
such an ill-behaved field. We note that $V_{R}$ is zero where the delta function representing the 
source density is non-vanishing. Hence, we evaluate the volume integral in Eq.(10) by deleting a small spherical volume 
element ${v}$ surrounding the point charge from the region of volume integration. 
We will take ${v} = 0$ at the end of calculation. The electrostatic energy is given by the following expression:

\be
\xi' = {{\epsilon_0} \over 2}{{\int_{(V - v)}}{({\vec{\nabla}}.{\vec{E}_{T}})}{V_{R}} d{\tau}}
\ee

\noindent{Here $v$ is a small spherical volume element centered at the point charge.
In the annular region of volume integration, we can replace $V_{R}$ by $V_{T}$
and apply the divergence theorem to the vector field $(V_{T}{{\vec{E}_{T}}})$ to evaluate
the integral. It is easy to find that ${\xi}'$ is vanishing and independent of $v$. This is 
consistent with the value of the energy obtained directly
from Eq.(10). We need not to consider any limiting
internal structure of the point charges to explain the 
self-energy of the point charges. This is considered in [14]
to explain the divergent expression for the electrostatic
self-energy. We will discuss this issue later.}

We now derive the expression for the electrostatic energy when we
have more than one point charge. We consider the situation with two
point charges. The electrostatic energy in terms of the potential is
given by the following expression:

\be
\xi = {{\epsilon_0} \over 2}{\int {({\vec{\nabla}}.{{\vec{E}_{T}}})}{V_{R}}
d{\tau}}
\ee

\noindent{In this case we no longer have ${V_{R}} = 0$ at the point charges.
Hence, we can not delete the infinitesimally small volume elements
containing the delta functions (the point charges) from our region
of integration. We find that the integral over the delta functions
give us the well-known expression for the energy of interaction. We
can delete two infinitesimal volume elements surrounding the point charges from our region of
volume integration provided we take into account this interaction
energy. The expression for the electrostatic energy with two point
charges is then given by the following expression:}

\be
\xi = {{\epsilon_0} \over 2}[{{\int_{({V - {v_1} - {v_2}})}}{({\vec{\nabla}}.{{\vec{E}_{T}}})}{V_{T}} d{\tau}}]
+ {{{{q_1}{q_2}}\over{4 \pi {\epsilon_0}}}{1 \over {R_{12}}}}
\ee

\noindent{Here, $({v_1}$ and ${v_2})$ are two infinitesimally small volume elements surrounding the point charges ${q_1}$ and ${q_2}$ respectively. One can show that the contribution
from the volume integration is zero. We should be careful about the surface terms. Thus we find that the electrostatic self energy
of a point charge is vanishing. We may consider a point charge to be the limiting situation of an small volume element whose total charge is always finite. One can then derive an expression for the total electrostatic energy
from Eq.(11). If we neglect the internal structure of such a source and corresponding electrostatic energy, we have an expression for the electrostatic energy which diverges inversely as the radius of the source tends to zero. This expression is similar to the conventional divergent expression of the electrostatic
self-energy of a point charge [14]. However, we should note that the situation with a sphere whose radius tends to zero is different from a sphere whose radius is exactly zero to represent a point. In the later situation, the Eq.(11) is no longer valid. This is related with the applicability of the Gauss' integral law to divergent fields. In this case the field $(V_{T}{{\vec{E}_{T}}})$ 
diverges as inverse cube of the distance from the point
charges and we can not apply the Gauss' integral law
as is done to obtain Eq.(11). This is discussed below
Eq.(11). The above discussions are significant when we
consider the Lagrangian description of Classical Electrodynamics. We
find that the applicability of Gauss integral law to
singular fields is a serious concern, when we try to construct the stress-tensor
in terms of fields with point particle sources. However, we can start from the Lagrangian density
in terms of the fields directly and evaluate the stress-tensor from the Lagrangian density. 
This will give singular expression for the self energy of a point charge.}

\newpage

\section*{3. Equivalence of the Schwarzschild and the
Kruskal-Szekers Coordinate System and Analytic Continuation}

The Schwarzschild space-time is a Lorentz signature, static
spherically symmetric solution of the Einstein equations when the
Ricci tensor vanishes. This solution describes the exterior geometry
of a static spherically symmetric star and has been used to verify
the predictions of general relativity for the Solar system.

A space-time is said to be static if there exits a space-like
hypersurface which is orthogonal to the orbits of the time-like
Killing vector field. A space-time is said to be spherically
symmetric if the space-like hypersurfaces contains $SO(3)$ as a
subgroup of the group of isometries. The orbit spheres of $SO(3)$
are isometric to the unit two sphere. These features together with
the condition of the asymptotic Newtonian limit give the well-known
Schwarzschild solution in the spherical polar coordinates [17]:

\be
ds^2 = -(1 - 2M/r) dt^2 + (1 - 2M/r)^{-1} dr^2 + r^2
[{d\theta}^2 + \sin^2{\theta}{d\theta}^2]
\ee

According to the Birkhoff's theorem [18] all spherically symmetric
solutions with $R_{ab} = 0$ are static and the Schwarzschild
space-time is the unique static spherically symmetric solution, up
to diffeomorphisims, of the Einstein equations with $R_{ab} = 0$.

The norm of the time-like Killing vector field and ${(\nabla r)}^a$
in the orthonormal coordinates vanishes and some of the metric
components are not well-behaved at $r = 2M$ in the Schwarzschild
coordinates. The proper acceleration of the constant $r$ observers
can be obtained from the geodesic equations in the Schwarzschild
coordinates. This acceleration, $a = {(1 - 2M/r)^{- 1/2}}{M/r^2}$,
is divergent at the horizon $(r = 2M)$.

The ill-behavednes of the Schwarzschild coordinates is not a
coordinate singularity like that of the spherical polar coordinate
system where the azimuthal angular coordinate $\phi$ become
ambiguous at the poles. All the ill-behavednes of the Schwarzschild
coordinates at the horizon originate from that of the space-time
metric. The curvature scalars calculated from the metric are
well-behaved at the horizon unlike $r = 0$ where the curvature
scalars diverge. For ordinary stars this metric singularity  at $r =
2M$ is irrelevant as it is inside the star and the Schwarzschild
solution is not valid in the matter filled interiors. However it is
well-known that sufficiently massive stars can undergo gravitational
collapse to form black holes and the metric singularity at the
horizon is important. Several coordinate systems had been introduced
to remove the metric singularity and to extend the Schwarzschild
space-time where the Schwarzschild coordinate system is referred to
covering a proper submanifold of the extended space-time. The metric
in these extended coordinate systems are well-defined every where
apart from the space-time singularity. The most well-known extension
is the Kruskal-Szekers coordinate system. In this article we perform
a comparative study of these two coordinate systems and show that
they are not diffeomorphically equivalent.

In this section we will follow the abstract index convention of Wald [17].
According to the theory of relativity if $\phi : M \rightarrow M$ is
diffeomorphism then $(M, g_{ab})$ and $(M, \phi^* g_{ab})$ represent
the same physical space-time. Let a coordinate system ${x^\mu}$
cover a neighborhood $U$ of a point $p$ and a coordinate system
${y^\nu}$ cover a neighborhood $V$ of the point $\phi(p)$. Now we
may use $\phi$ to define a new coordinate system ${{x'}^\mu}$ in a
neighborhood $O = {\phi^{-1}}[V]$ by setting ${{x'}^\mu} = y^\mu
[\phi (q)]$ for $q$ belonging to $O$. We may then take the point of
view as $\phi$ leaving $p$ and all tensors at $p$ unchanged but
inducing the coordinate transformation ${x^\mu} \rightarrow
{{x'}^\mu}$. For $\phi$ to be a diffeomorphism ${{\partial
{{x'}^\mu}} \over{\partial x^\nu}}$ should be non-singular [17,18].
According to this point of view two coordinate system covering a
space-time can be taken to be equivalent if the corresponding
transformation coefficients are not singular in their common domain
of definition otherwise an arbitrary smooth function defined in one
coordinate system may not remain smooth in the other coordinate
system.

To extend the Schwarzschild coordinate system one considers
the two dimensional $r-t$ part:

\be
ds^2 = -(1 - 2M/r) dt^2 + (1 - 2M/r)^{-1} dr^2
\ee

The Regge-Wheeler coordinate system is defined through
the null-geodesics and is given by:

\be
r_* = r + 2M ln (r/2M - 1)
\ee

in this coordinate $r \rightarrow 2M$ corresponds to
$r_* \rightarrow -\infty$. The null coordinates are defined
as:

\be
u = t - r_* , ~~v = t + r_*
\ee

A regular metric is obtained through the following transformation,

\be
U = -e^{-u/4M} , ~~V = e^{v/4M}
\ee

The metric in these coordinates becomes:

\be
ds^2 = -{{32M^3 e^{-r/2M}} \over {r}}dUdV
\ee

As there is no longer a coordinate singularity at $r = 2M$ (i.e at
$U = 0$ or $V = 0$) one extends the Schwarzschild solution by
allowing $U,V$ to take all possible values. However the
transformation coefficients $dU/dr = -d[{(r/2M - 1)^{1/2} e^{-{(t -
r) \over{4M}}}}]/dr$ and $dV/dr = d[{(r/2M - 1)^{1/2} e^{{(t + r)
\over{4M}}}}]/dr$ are singular at $r = 2M$ and the extension is not
diffeomorphically equivalent. Consequently as discussed at the
beginning of this section the Schwarzschild coordinate system and
the $(U,V)$ coordinate system do not represent physically the same
space-time manifold. Consequently, according to Birkoff's theorem,
the space-time represented by the $(U,V,\theta,\phi)$ coordinate
system is not a solution of the Einstein equations for a spherically
symmetric black hole.

Similar discussions are valid for the Kruskal-Szekers coordinate
transformations which are obtained through the following
transformations:

\be
T = (U + V)/2 , ~~X = (V - U)/2
\ee

and the metric becomes,

\be
ds^2 = {{32M^3 e^{-r/2M}} \over {r}}(-dT^2 + dX^2)
+ r^2{({d\theta}^2 + \sin^2{\theta}{d\phi}^2)}
\ee

The relation between the $(T,X)$ and the $(t,r)$ coordinates are
well known and in the physical regions of interests are given by
[8],

\be
X = (r/2M - 1)^{1/2} e^{r/4M} {\cosh(t/4M)}
\ee

\be
T = (r/2M - 1)^{1/2} e^{r/4M} {\sinh(t/4M)}
\ee

valid for $r ~> ~2M$, and

\be
T = (1 - r/2M)^{1/2} e^{r/4M} {\cosh(t/4M)}
\ee

\be
X = (1 - r/2M)^{1/2} e^{r/4M} {\sinh(t/4M)}
\ee

valid for $r ~< ~2M$.

Again the transformation coefficients are not defined on the horizon
and the Kruskal-Szekers coordinates do not give a proper
diffeomorphic extension of the Schwarzschild coordinate system.
Hence the Kruskal-Szekeres coordinates is not a solution of the
Einsteins equations for a spherically symmetric black hole

The Kruskal-Szekers coordinate system had been introduced to
eliminate a particular singular function (the metric components) in
the Schwarzschild coordinate system through a singular coordinate
transformation. This does not ensure that all singular tensors can
be made regular in the new coordinate system and also tensors which
are regular in the $(t,r)$ coordinates can become singular in the
$(T,R)$ coordinates. To illustrate these features we consider the
implicit relations between the two coordinate systems [17]:

\be
(r/2M - 1)e^{r/2M} = X^2 - T^2
\ee

\be
{t \over 2M} = ln({{T + X} \over {X - T}})
\ee

The horizon in this coordinates are defined as $X = \pm T$.

Firstly the proper acceleration of the curves in Kruskal-Szecker's
coordinate system which correspond to the constant $r$ observers in
the Schwarzschild coordinate system is given by $a = (X^2 -
T^2)^{-{1/2}}[e^{r/2M}{M/r^2}]$. This is also divergent on the
horizon.

Secondly we consider the vector $({{dR} \over {ds}})^a$, $R^{'a}$,
the proper rate of change of the curvature scalar $R$ obtained from
${(dR)^a}$ and the proper distance $ds$ [i.e, the vector $({{dR}
\over {ds}})({{\partial} \over {\partial{\bf r}}})$].
The norm of this vector in the Schwarzschild coordinate
system is $(dR/dr)^2$ and is finite on the horizon. Whereas the
corresponding quantity in the $(T,X)$ coordinates can be
obtained from the following relations
[apart from normalizing factors: $({{\partial X} \over {\partial s}}),
({{\partial T} \over {\partial s}}) =
{[{{re^{r/2M}} \over {32M^3}}]^{1/2}}:$

\be
{dR \over dX} = {{\partial R}\over {\partial r}}
{{\partial r}\over {\partial X}} , ~~{dR \over dT} =
{{\partial R}\over {\partial r}}
{{\partial r}\over {\partial T}}
\ee

and from Eq.(65),

\be
{{\partial r}\over {\partial X}} = {{8 M^2 X e^{-{r/2M}}}
\over{r}},
~~{{\partial r}\over {\partial T}} = -{{8 M^2 T e^{-{r/2M}}}
\over{r}}
\ee

and we have $|{R^{'a}}_{KS}|^2 = {{64 M^4 e^{- r/M}} \over
{r^2}}{({{\partial R}\over {\partial r}})^2} [X^2 - T^2] = 0$ on the
horizon although the $r$ -dependent multiplying factor in front of
the Kruskal-Szecker's metric is finite at $r = 2M$.

The unit space-like normal vector to the $r = constant$ surfaces,
which can be defined apart from $r = 0$, $k^a =
{({{dr}\over{ds}})^a}$ has unit norm ($k^a k_a = 1$) on $r = 2M$
although $k^a \rightarrow 0$ as $r \rightarrow 2M$  which for an
outside observer ($r ~> ~2M$) may be interpreted as nothing can
propagate radially outward at $r = 2M$.

For two metric spaces the definitions of continuity
is as follows [11]:

Let $(S, d_S)$ and $(T, d_T)$ be metric spaces and let $f: S
\rightarrow T$ be a function from $S$ to $T$. The function $f$ is
said to be continuous at a point $p$ in $S$ if for every
infinitesimal $\epsilon > 0$ there is an infinitesimal $\delta > 0$
such that

\be
{d_T}{[f(x),f(p)]} < \epsilon, ~~whenever ~{d_S}{[x,p]} < \delta.
\ee

If $f$ is continuous at every point of $S$ then $f$ is
continuous on $S$.

The definition is in accordance with the intuitive idea that points
close to $p$ are mapped by $f$ into points closed to $f(p)$. From
Eqs.(27,28) we have,

\be
|dt|_{Sch} = {{X} \over {(X^2 - T^2)^{1/2}}}{|dT|_{KS}},
~~|dt|_{Sch} = -{{T} \over {(X^2 - T^2)^{1/2}}}{|dX|_{KS}}
\ee

and,

\be
|dr|_{Sch} = {{X} \over {(X^2 - T^2)^{1/2}}}{|dX|_{KS}},
~~|dr|_{Sch} = -{{T} \over {(X^2 - T^2)^{1/2}}}{|dT|_{KS}}
\ee

where $|{~~}|$ denotes the proper elements (for proper distances we have to consider 'i') 
in the respective coordinate systems
and we find that the coordinate transformation, $(t,r) \rightarrow
(T,X)$ is not continuous on the horizon as the multiplicative
factors diverge on the horizon $(X = \pm T)$. Consequently the
coordinate transformation $(t,r) \rightarrow (T,X)$ is not a
homeomorphism and the two coordinate systems do not topologically
represent the same space-time manifolds [11,19]. Hence we show that
that the Kruskal-Szekers coordinate system is not a proper extension
of the Schwarzschild coordinate system and it is not a solution of
the Einsteins equation for spherically symmetric black hole. We
conclude this discussion with the following note:

For any coordinate system we have,

\be
{g'}_{\mu \nu} = {{\partial {x^ \rho}}\over {\partial {{x'}^
\mu}}} {{\partial {x^ \sigma}}\over {\partial {{x'}^ \nu}}}
{({g_{Sch.}})_{\rho \sigma}}
\ee

Consequently it is not possible to find a coordinate system with a
regular ${g'}_{\mu \nu}$ without absorbing the singularities of
${({g_{Sch.}})_{\rho \sigma}}$ at $r = 2M$ into the transformation
coefficients ${{\partial {x^ \rho}}\over {\partial {{x'}^ \mu}}}$ at
$r = 2M$ i.e, without breaking the diffeomorphic equivalence of the
two coordinate systems. Thus, as also discussed in the preceding
sections, the Kruskal-Szekeres coordinate system with a regular
metric at the horizon may not be diffeomorphically equivalent to the
Schwarzschild coordinate system. We can improve the situation using
analytic continuation [18].

\newpage

\section*{4.1 Length contraction and Michelson-Morley experiment}

We consider the Michelson-Morley experiment from the space-fixed
frame. The partially silvered mirror that splits light into two perpendicular
directions bends in the direction opposite to the velocity
of the apparatus. In a thought experiment, we can use a narrow beam of light, and
for sufficiently high velocity of the apparatus, there can be no interference fringe
with respect to the space-fixed observer. This leads to a paradoxical situation
since the interference fringe will always be present with respect to the rest frame
of the apparatus. However, we should also consider the laws of reflection of light from a moving mirror
with relativistic velocity.

\section*{4.2 Comments on Hydrodynamics}

In this article we will review the laws of fluid dynamics. Our
discussions will be based on mainly that of chapter 40, 41 of The
Feynman Lectures on Physics, Vol.2 [16].

The dynamics of dry water is governed by Eq.(40.6) [16]:

\be
{\partial{\vec{v}}\over{\partial t}}
+ ({\vec{v}}.{\vec{\nabla}}){\vec v} =
-{{{\vec{\nabla}}p}\over{\rho}} - {{\vec{\nabla}}{\phi}}
\ee

or using a vector analysis identity to the second
term of the above equation:

\be
{\partial{\vec{v}}\over{\partial t}}
+ {({\vec{\nabla} \times {\vec{v}}}) \times {\vec{v}}}
= -{{{\vec{\nabla}}p}\over{\rho}} - {{\vec{\nabla}}{\phi}}
\ee

where $\vec{v}$ is the velocity of an fluid element
for which such laws can be applicable, $p$ is the
fluid pressure and $\phi$ is the potential per
unit mass for any potential force present.
We can derive some important laws from Eq.(35).
The first one is the equation for vorticity
$(\Omega = {\vec{\nabla} \times {\vec{v}}})$
and is obtained by taking curl of Eq.(36):

\be
{\partial{\vec{\Omega}}\over{\partial t}}
+ {{\vec{\nabla}} \times ({\vec{\Omega}} \times {\vec{v}})}
= 0
\ee

The second one is Bernoulli's theorems (40.12) and (40.14) [16]:

\be
{\vec{v}}.{\vec{\nabla}}({p\over \rho} + \phi + {1\over 2}{v^2}) = 0
\ee

i.e,

\be
{p\over \rho} + \phi + {1\over 2}{v^2} =
const {~} (along {~} streamlines)
\ee

valid for
steady flow and

\be
{p\over \rho} + \phi + {1\over 2}{v^2} =
const {~} (everywhere)
\ee

valid for steady and irrotational flow.

However in all these equations the variation of the fluid density,
$\rho$, is not considered while in deriving Eq.(40.17) [16] the
variation of fluid density is not properly taken into account. The
consideration for variation of the density of a
nearly-incompressible fluid may become important through the facts
that when unconstrained the shape of a fluid can be changed almost
freely and sparsed away and through the facts that layers of fluids
can be very easily spread or detached away although these properties
vary from fluid to fluid. These features together with the local
version of the conservation of mass law (assuming that there is no
local source or sink in the region of interest):

\be
{\vec{\nabla}}.(\rho \vec{v}) = 0
\ee

indicate that we should consider the possibility for variation of
$\rho$ properly as we will illustrate later that some ideal models
can cause a finite variation of $\rho$ and in reality the
description of the motion should be changed. While the divergence of
$\vec{v}$ may become important in cases like Couette flow where the
centrifugal forces imposes a finite and may even be large divergence
of $\vec{v}$.

We can derive a proper version of Eq.(35) by applying
Newtons second law to the fluid momentum per unit volume
and we have:

\be
{\partial{(\rho\vec{v})}\over{\partial t}}
+ [{\vec{v}}.{\vec{\nabla}}]{(\rho\vec v)} =
-{{\vec{\nabla}}p} - {{\vec{\nabla}}{(\rho\phi)}}
\ee

This equation is in general a non-linear coupled
[through Eq.(26)] partial differential
equation for $\vec{v}$.

Bernoulli's theorems for
fluid dynamics can only be established when
$\rho$ is constant :

\be
{\vec{v}}.{\vec{\nabla}}[p + (\phi \rho) +
{1\over 2}{(\rho v^2)}] = 0
\ee

i.e,

\be
p + (\phi \rho) +
{1\over 2}{(\rho v^2)} =
const {~} (along {~} streamlines)
\ee

valid for
steady flow and

\be
p + (\phi \rho) + {1\over 2}{(\rho v^2)} =
const {~} (everywhere)
\ee

valid for steady and irrotational flow.

In general, when $\rho$ is varying,
only the first of the Bernoulli's theorems :

\be
{\vec{v}}.{\vec{\nabla}}[p + (\phi \rho) +
{1\over 2}{(\rho v^2)}] = 0
\ee

remains to be valid provided $({\vec{v}}.{{\vec{\nabla}}\rho})$ is
vanishing or is approximately valid if
$|({\vec{v}}.{{\vec{\nabla}}\rho}){\vec{v}}|$ is negligible compared
to the other terms in Eq.(42). To illustrate the significance of
these comments, let us consider the ideal model to calculate the
efflux-coefficient, fig. 40-7 [16]. After that the contraction of
the cross-section of the emerging jet has stopped we have, from the
conservation of mass law, the following equation for $\rho {v}$ at
two vertical points:

\be
{\rho}_1{v_1} = {\rho}_2{v_2}
\ee

In this case pressure is the atmospheric pressure
and remains the same throughout the flow and
thus even for the flow of a nearly-incompressible
fluid $\rho$ can vary as $v$ changes with height. In reality
the flow usually gets sparsed away after a distance
which varies for different flows.

The viscous flow of a fluid
is governed by the following two laws which
are obtained from Eq.(42) and Eq.(41.15),[16]:

\be
{\partial{(\rho\vec{v})}\over{\partial t}}
+ [{\vec{v}}.{\vec{\nabla}}]{(\rho\vec v)} =
-{{\vec{\nabla}}p} - {{\vec{\nabla}}{(\rho\phi)}}
+ {\eta}{{{\nabla}^2}\vec{v}} +
({\eta} + {\eta}'){{\vec{\nabla}}({\vec{\nabla}}.{\vec{v}})}
\ee

\be
{\vec{\nabla}}.(\rho \vec{v}) = 0
\ee

supplemented by proper boundary conditions. To illustrate the
significance of the boundary conditions we can consider the change
of the shape of the surface of water in a bucket when the bucket is
given a steady rotational motion about it's axis. The surface of the
water become paraboloidal when the bucket is rotating. This shape
can not be obtained without a vertical component of fluid velocity
along the bucket surface for a finite duration although the bucket
surface only have an angular velocity.

In the above equations
$\eta$ is the ``first coefficient of viscosity''
or the ``shear viscosity coefficient'' and
${\eta}'$ is the ``second coefficient of viscosity''.
This equation is significant in the sense
that this equation, not Eq.(41.16) [16], is the
equation which contains all the terms relevant
to describe the dynamics of  viscous fluids,
both nearly-incompressible and compressible.
For compressible fluids $\rho$ will also
depend on pressure, $p({\vec{r}})$.
We can modify this equation only through
varying the nature of the viscous force.

The equation for vorticity is given by:

\ba
{\partial{\vec{\Omega}}\over{\partial t}}
+ {{\vec{\nabla}} \times ({\vec{\Omega}} \times {\vec{v}})}
- {{\vec{\Omega}}({\vec{\nabla}}.{\vec{v}})}
+ {{\vec{v}} \times {{\vec{\nabla}}({\vec{\nabla}}.\vec{v})}}
= {{\eta \over \rho}({{\nabla}^2}{\vec{\Omega}})}
{~~~~~~~~~~~~~} & {~} & \nonumber\\
- {{({{\vec{\nabla}}p}) \times ({{\vec{\nabla}}\rho})}
\over {{\rho}^2}}
- {{[{{\vec{\nabla}}{(\rho\phi)}] \times ({{\vec{\nabla}}\rho})}
\over {{\rho}^2}}}
+ {{{\eta}({{{\nabla}^2}{\vec v}}) \times ({{\vec{\nabla}}\rho})}
\over {{\rho}^2}}
+ {{\eta + {\eta}'}\over {{\rho}^2}}
{{\vec{\nabla}({\vec{\nabla}}.{\vec{v}})} \times ({{\vec{\nabla}}\rho})}
\ea

We can obtain an equation similar to Eq.(41.17) [16]
describing the motion of a viscous fluid past a cylinder
provided we can neglect the terms involving ${{\vec{\nabla}}\rho}$
and it is given by:

\be
{\partial{\vec{\Omega}}\over{\partial t}}
+ {{\vec{\nabla}} \times ({\vec{\Omega}} \times {\vec{v}})}
- {{\vec{\Omega}}({\vec{\nabla}}.{\vec{v}})}
+ {{\vec{v}} \times {{\vec{\nabla}}({\vec{\nabla}}.\vec{v})}}
= {{\eta \over \rho}({{\nabla}^2}{\vec{\Omega}})}
\ee

Following the procedure
in section 41-3,[16] we can rescale the variables to obtain an equation
which has Reynolds number $(R)$ as the only free
parameter :

\be
{\partial{\vec{\omega}}\over{\partial t'}}
+ {{\vec{\nabla}'} \times ({\vec{\omega}} \times {\vec{u}})}
- {{\vec{\omega}}({\vec{\nabla}'}.{\vec{u}})}
+ {{\vec{u}} \times {{\vec{\nabla}'}({\vec{\nabla}'}.\vec{u})}}
= {{1 \over R}({{{\nabla}'}^2}{\vec{\omega}})}
\ee

where the prime describe the scaled variables,
$\vec{u}$ is the scaled velocity
and $R$ is given by the usual expression,
$R = {{\eta \over \rho}V D}$.

To conclude in this section we have derived the exact equation
describing fluid dynamics. We considered the motion of both
non-viscous and viscous fluids. We proved that in both the cases
there are terms which are neglected in the conventional theory but
may become significant in some ideal model and in reality the
description of motion is changed. Some of these terms even change
the dynamical laws of viscous fluid motions by violating the
conventional theory established in term of the Reynold number and
these terms are significant for the dynamics of compressible fluids
like air.

\section*{4.3 Electrons, holes and quasi-particles:} In a p-type semiconductor the motion of holes are out of the movements of
the valence band or the acceptor level electrons. The electrons experience forces (qvB) in the same direction as that of
the holes and in terms of the motion of the electrons the polarity of the Hall voltage should be opposite.
The Hall voltage is explained by considering the holes. What is the motion of the electrons to give the Hall voltage?

\section{{References}}

[1] K. Ghosh; International Journal of Pure and Applied Mathematics,
    \textbf{76}, No.2, pp.251 - 260, (2012).

[2] James R. Munkres; Topology A First Course, (Prentice-Hall of India Private Limited, 1994).

[3] Paul J. Cohen; Set Theory and the Continuum Hypothesis, (Dover Publications, Inc., 1994).

[4] R. Dedekind; Irrational Numbers, The World of Mathematics: Vol.\textbf{1}, Editor: James Newman,
    (Dover Publications, Inc., 1956).
		
[5] R. Courant and F. John; Introduction to Calculus and Analysis: Vol.\textbf{1}, (Springer-Verlag New York Inc., 1989).		
		
[6] L. Carroll; Continuity, The World of Mathematics: Vol.\textbf{4}, Editor: James Newman,
    (Dover Publications, Inc., 1956).

[7] J. Polchinski; String Theory: Vol.\textbf{I}, (Cambridge University Press, 1998).

[8] C. W. Misner, K. S. Thorne and J. A. Wheeler; Gravitation (W.H. Freeman and company, New York, 1970).

[9] John G. Hocking and Gail S. Young; Topology (Dover Publications, Inc., New York, 1961). 

[10] Tom M. Apostol; Mathematical Analysis (Narosa Publishing House, 1992).

[11] S. Kobayashi and K. Nomizu; Foundations of Differential Geometry: Vol. I 
(Wiley Classics Library, 1991).  

[12] K. Ghosh; Physics of the Dark Universe, \textbf{26}, 100403 (2019).

[13] K. Ghosh; International Journal of Pure and Applied Mathematics,
    Volume: \textbf{76}, No.2, pp.207 (2012); Academic Publishers.

[14] D. J. Griffiths; Introduction To Electrodynamics (Prentice-Hall of India, 1989). 

[15] J. D. Jackson; Classical Electrodynamics (Wiley Eastern Limited, 1990).

[16] R. P. Feynman, R. B. Leighton and M. Sands; The Feynman Lectures
     on Physics Vol.II (Narosa Publishing House, 1986).

[17] R. M. Wald; General Relativity (The University of Chicago Press, Chicago and London, 1984).

[18] S. W. Hawking and G. F. R. Ellis; The Large Scale Structure
     of Space-Time (Cambridge University Press, 1973).
		
[19] M. Nakahara; Geometry, Topology And Physics (Adam Hilger, Bristol And New York, 1990).
		
[20] K. Ghosh, Prog. Theor. Exp. Phys.{\bf 2016}, 093E03.

[21] C. Kittel; Introduction to Solid State Physics (Wiley Eastern Limited, New Delhi, 1994).

[22] K. Huang; Statistical Mechanics (Wiley-India Ltd., New Delhi, 2003).

\end{document}